\documentclass[twocolumn]{aastex63}
\usepackage{graphicx}
\usepackage{float}
\usepackage[T1]{fontenc}
\usepackage{aecompl}
\usepackage[]{amsmath}
\usepackage{color}
\usepackage{xspace}
\usepackage{soul}
\usepackage{booktabs} 
\usepackage{array} 
\usepackage{multirow} 
\usepackage{orcidlink}
\usepackage{booktabs,tabularx}  
\usepackage{amsmath}
\usepackage{rotating}   
\usepackage{multirow} 
\usepackage{lineno}
\usepackage{gensymb}

\graphicspath{{figures/}}

\submitjournal{ApJ}

\shorttitle{DESI Y3 stellar halo profile and shape}
\shortauthors{Li et al.}

\graphicspath{{./}{figures/}}

\begin{document}

\title{The Milky Way stellar halo is twisted and doubly broken: insights from DESI DR2 Milky Way Survey observation}

\author[0000-0002-6469-8263]{Songting Li}
\thanks{songtingli@sjtu.edu.cn}
\affiliation{Department of Astronomy, School of Physics and Astronomy, and Key Laboratory for Particle Astrophysics and Cosmology (MOE)/Shanghai Key Laboratory for Particle Physics and Cosmology, Shanghai Jiao Tong University, Shanghai 200240, People's Republic of China}
\affiliation{Tsung-Dao Lee Institute, Shanghai Jiao Tong University, Shanghai, 201210, China}
\affiliation{State Key Laboratory of Dark Matter Physics, School of Physics and Astronomy, Shanghai Jiao Tong University, Shanghai 200240, China}
\author[0000-0002-5762-7571]{Wenting Wang}
\thanks{corresponding author: wenting.wang@sjtu.edu.cn}
\affiliation{Department of Astronomy, School of Physics and Astronomy, and Key Laboratory for Particle Astrophysics and Cosmology (MOE)/Shanghai Key Laboratory for Particle Physics and Cosmology, Shanghai Jiao Tong University, Shanghai 200240, People's Republic of China}
\affiliation{State Key Laboratory of Dark Matter Physics, School of Physics and Astronomy, Shanghai Jiao Tong University, Shanghai 200240, China}
\author{Sergey E. Koposov}
\thanks{Sergey.Koposov@ed.ac.uk}
\affiliation{Institute for Astronomy, University of Edinburgh, Royal Observatory, Blackford Hill, Edinburgh EH9 3HJ, UK}
\affiliation{Institute of Astronomy, University of Cambridge, Madingley Road, Cambridge CB3 0HA, UK}
\author[0000-0002-7662-5475]{Jo\~{a}o A. S. Amarante}
\affiliation{Department of Astronomy, School of Physics and Astronomy, and Shanghai Key Laboratory for Particle Physics and Cosmology, Shanghai Jiao Tong University, Shanghai 200240, People's Republic of China}
\affiliation{State Key Laboratory of Dark Matter Physics, School of Physics and Astronomy, Shanghai Jiao Tong University, Shanghai 200240, China}
\author{Alis J. Deason}
\affiliation{Institute for Computational Cosmology, Department of Physics, Durham University, South Road, Durham DH1 3LE, UK}
\author[0000-0002-7393-3595]{Nathan R. Sandford}
\affiliation{David A. Dunlap Department of Astronomy \& Astrophysics, University of Toronto, 50 St George Street, Toronto ON M5S 3H4, Canada}
\author{Ting S. Li}
\affiliation{David A. Dunlap Department of Astronomy \& Astrophysics, University of Toronto, 50 St George Street, Toronto ON M5S 3H4, Canada}
\author[0000-0003-0105-9576]{Gustavo E. Medina}
\affiliation{David A. Dunlap Department of Astronomy \& Astrophysics, University of Toronto, 50 St George Street, Toronto ON M5S 3H4, Canada}
\author{Jiaxin Han}
\affiliation{Department of Astronomy, School of Physics and Astronomy, and Key Laboratory for Particle Astrophysics and Cosmology (MOE)/Shanghai Key Laboratory for Particle Physics and Cosmology, Shanghai Jiao Tong University, Shanghai 200240, People's Republic of China}
\affiliation{State Key Laboratory of Dark Matter Physics, School of Physics and Astronomy, Shanghai Jiao Tong University, Shanghai 200240, China}
\author{Monica Valluri}
\affiliation{Department of Astronomy and Astrophysics, University of Michigan, Ann Arbor, MI, USA}
\author{Oleg~Y.~Gnedin}
\affiliation{Department of Astronomy, University of Michigan, Ann Arbor, MI 48109, USA}
\author{Namitha Kizhuprakkat}
\affiliation{Institute of Astronomy and Department of Physics, National Tsing Hua University, Hsinchu 30013, Taiwan}
\affiliation{Center for Informatics and Computation in Astronomy, National Tsing Hua University, Hsinchu 30013, Taiwan}
\author{Andrew P. Cooper}
\affiliation{Institute of Astronomy and Department of Physics, National Tsing Hua University, Hsinchu 30013, Taiwan}
\affiliation{Center for Informatics and Computation in Astronomy, National Tsing Hua University, Hsinchu 30013, Taiwan}
\author[0000-0002-0740-1507]{Leandro {Beraldo e Silva}}
\affiliation{Observatório Nacional, Rio de Janeiro - RJ, 20921-400, Brasil}
\author{Carlos Frenk}
\affiliation{Institute for Computational Cosmology, Department of Physics, Durham University, South Road, Durham DH1 3LE, UK}
\author[0000-0002-7667-0081]{Raymond G. Carlberg}
\affiliation{David A. Dunlap Department of Astronomy \& Astrophysics, University of Toronto, 50 St George Street, Toronto ON M5S 3H4, Canada}
\author[0000-0002-2527-8899]{Mika Lambert}
\affiliation{Department of Astronomy \& Astrophysics, University of California, Santa Cruz, 1156 High Street, Santa Cruz, CA 95064, USA}
\author[0000-0002-4900-2088]{Tian Qiu}
\affiliation{Department of Astronomy, School of Physics and Astronomy, and Key Laboratory for Particle Astrophysics and Cosmology (MOE)/Shanghai Key Laboratory for Particle Physics and Cosmology, Shanghai Jiao Tong University, Shanghai 200240, People's Republic of China}
\affiliation{State Key Laboratory of Dark Matter Physics, School of Physics and Astronomy, Shanghai Jiao Tong University, Shanghai 200240, China}
\author{Jessica Nicole Aguilar}
\affiliation{Lawrence Berkeley National Laboratory, 1 Cyclotron Road, Berkeley, CA 94720, USA}
\author[0000-0001-6098-7247]{Steven Ahlen}
\affiliation{Department of Physics, Boston University, 590 Commonwealth Avenue, Boston, MA 02215 USA}
% \author[0000-0002-0084-572X]{Carlos Allende Prieto}
% \affiliation{Instituto de Astrof\'{\i}sica de Canarias, C/ V\'{\i}a L\'{a}ctea, s/n, E-38205 La Laguna, Tenerife, Spain}
% \affiliation{Departamento de Astrof\'{\i}sica, Universidad de La Laguna (ULL), E-38206, La Laguna, Tenerife, Spain}
\author[0000-0001-9712-0006]{Davide Bianchi}
\affiliation{Dipartimento di Fisica ``Aldo Pontremoli'', Universit\`a degli Studi di Milano, Via Celoria 16, I-20133 Milano, Italy}
\affiliation{INAF-Osservatorio Astronomico di Brera, Via Brera 28, 20122 Milano, Italy}
\author{David Brooks}
\affiliation{Department of Physics \& Astronomy, University College London, Gower Street, London, WC1E 6BT, UK}
\author{Todd Claybaugh}
\affiliation{Lawrence Berkeley National Laboratory, 1 Cyclotron Road, Berkeley, CA 94720, USA}
\author[0000-0002-1769-1640]{Axel de la Macorra}
\affiliation{Instituto de F\'{\i}sica, Universidad Nacional Aut\'{o}noma de M\'{e}xico, Circuito de la Investigaci\'{o}n Cient\'{\i}fica, Ciudad Universitaria, Cd. de M\'{e}xico C.~P.~04510, M\'{e}xico}
\author{Peter Doel}
\affiliation{Department of Physics \& Astronomy, University College London, Gower Street, London, WC1E 6BT, UK}
% \author[0000-0002-3033-7312]{Andreu Font-Ribera}
% \affiliation{Institut de F\'{i}sica d’Altes Energies (IFAE), The Barcelona Institute of Science and Technology, Edifici Cn, Campus UAB, 08193, Bellaterra (Barcelona), Spain}
\author[0000-0002-2890-3725]{Jaime E. Forero-Romero}
\affiliation{Departamento de F\'isica, Universidad de los Andes, Cra. 1 No. 18A-10, Edificio Ip, CP 111711, Bogot\'a, Colombia}
\affiliation{Observatorio Astron\'omico, Universidad de los Andes, Cra. 1 No. 18A-10, Edificio H, CP 111711 Bogot\'a, Colombia}
\author[0000-0001-9632-0815]{Enrique Gaztañaga}
\affiliation{Institut d'Estudis Espacials de Catalunya (IEEC), c/ Esteve Terradas 1, Edifici RDIT, Campus PMT-UPC, 08860 Castelldefels, Spain}
\affiliation{Institute of Cosmology and Gravitation, University of Portsmouth, Dennis Sciama Building, Portsmouth, PO1 3FX, UK}
\author[0000-0003-3142-233X]{Satya Gontcho A Gontcho}
\affiliation{Lawrence Berkeley National Laboratory, 1 Cyclotron Road, Berkeley, CA 94720, USA}
\affiliation{University of Virginia, Department of Astronomy, Charlottesville, VA 22904, USA}
\author{Gaston Gutierrez}
\affiliation{Fermi National Accelerator Laboratory, PO Box 500, Batavia, IL 60510, USA}
% \author[0000-0003-0201-5241]{Jorge Jimenez}
% \affiliation{Institut de F\'{i}sica d’Altes Energies (IFAE), The Barcelona Institute of Science and Technology, Edifici Cn, Campus UAB, 08193, Bellaterra (Barcelona), Spain}
\author[0000-0003-0201-5241]{Dick Joyce}
\affiliation{NSF NOIRLab, 950 N. Cherry Ave., Tucson, AZ 85719, USA}
\author{Robert Kehoe}
\affiliation{Department of Physics, Southern Methodist University, 3215 Daniel Avenue, Dallas, TX 75275, USA}
\author[0000-0001-6356-7424]{Anthony Kremin}
\affiliation{Lawrence Berkeley National Laboratory, 1 Cyclotron Road, Berkeley, CA 94720, USA}
\author[0000-0002-6731-9329]{Claire Lamman}
\affiliation{The Ohio State University, Columbus, 43210 OH, USA}
\author[0000-0003-1838-8528]{Martin Landriau}
\affiliation{Lawrence Berkeley National Laboratory, 1 Cyclotron Road, Berkeley, CA 94720, USA}
\author[0000-0001-7178-8868]{Laurent Le Guillou}
\affiliation{Sorbonne Universit\'{e}, CNRS/IN2P3, Laboratoire de Physique Nucl\'{e}aire et de Hautes Energies (LPNHE), FR-75005 Paris, France}
% \author[0000-0002-1125-7384]{Aaron Meisner}
% \affiliation{NSF NOIRLab, 950 N. Cherry Ave., Tucson, AZ 85719, USA}
\author{Ramon Miquel}
\affiliation{Institut de F\'{i}sica d’Altes Energies (IFAE), The Barcelona Institute of Science and Technology, Edifici Cn, Campus UAB, 08193, Bellaterra (Barcelona), Spain}
\affiliation{Instituci\'{o} Catalana de Recerca i Estudis Avan\c{c}ats, Passeig de Llu\'{\i}s Companys, 23, 08010 Barcelona, Spain}
% \author[0000-0001-9070-3102]{Seshadri Nadathur}
% \affiliation{Institute of Cosmology and Gravitation, University of Portsmouth, Dennis Sciama Building, Portsmouth, PO1 3FX, UK}
\author[0000-0002-0644-5727]{Will Percival}
\affiliation{Department of Physics and Astronomy, University of Waterloo, 200 University Ave W, Waterloo, ON N2L 3G1, Canada}
\affiliation{Perimeter Institute for Theoretical Physics, 31 Caroline St. North, Waterloo, ON N2L 2Y5, Canada}
\affiliation{Waterloo Centre for Astrophysics, University of Waterloo, 200 University Ave W, Waterloo, ON N2L 3G1, Canada}
% \author{Claire Poppett}
% \affiliation{University of California, Berkeley, 110 Sproul Hall \#5800 Berkeley, CA 94720, USA}
% \affiliation{Space Sciences Laboratory, University of California, Berkeley, 7 Gauss Way, Berkeley, CA 94720, USA}
% \affiliation{Lawrence Berkeley National Laboratory, 1 Cyclotron Road, Berkeley, CA 94720, USA}
\author[0000-0001-7145-8674]{Francisco Prada}
\affiliation{Instituto de Astrof\'{i}sica de Andaluc\'{i}a (CSIC), Glorieta de la Astronom\'{i}a, s/n, E-18008 Granada, Spain}
\author[0000-0001-6979-0125]{Ignasi Pérez-Ràfols}
\affiliation{Departament de F\'isica, EEBE, Universitat Polit\`ecnica de Catalunya, c/Eduard Maristany 10, 08930 Barcelona, Spain}
\author{Graziano Rossi}
\affiliation{Department of Physics and Astronomy, Sejong University, 209 Neungdong-ro, Gwangjin-gu, Seoul 05006, Republic of Korea}
\author[0000-0002-9646-8198]{Eusebio Sanchez}
\affiliation{CIEMAT, Avenida Complutense 40, E-28040 Madrid, Spain}
\author{David Schlegel}
\affiliation{Lawrence Berkeley National Laboratory, 1 Cyclotron Road, Berkeley, CA 94720, USA}
\author[0000-0003-3449-8583]{Ray Sharples}
\affiliation{Centre for Advanced Instrumentation, Department of Physics, Durham University, South Road, Durham DH1 3LE, UK}
\affiliation{Institute for Computational Cosmology, Department of Physics, Durham University, South Road, Durham DH1 3LE, UK}
\author[0000-0002-3461-0320]{Joseph Harry Silber}
\affiliation{Lawrence Berkeley National Laboratory, 1 Cyclotron Road, Berkeley, CA 94720, USA}
\author{David Sprayberry}
\affiliation{NSF NOIRLab, 950 N. Cherry Ave., Tucson, AZ 85719, USA}
\author[0000-0003-1704-0781]{Gregory Tarlé}
\affiliation{University of Michigan, 500 S. State Street, Ann Arbor, MI 48109, USA}
\author{Benjamin Alan Weaver}
\affiliation{NSF NOIRLab, 950 N. Cherry Ave., Tucson, AZ 85719, USA}
\author[0000-0002-6684-3997]{Hu Zou}
\affiliation{National Astronomical Observatories, Chinese Academy of Sciences, A20 Datun Road, Chaoyang District, Beijing, 100101, P.~R.~China}

\clearpage

\begin{abstract}
Using K giants from the second data release (DR2) of the Dark Energy Spectroscopic Instrument (DESI) Milky Way (MW) Survey, we measure the shape, orientation, radial profile, and density anisotropies of the MW stellar halo over 8~kpc$<r_\mathrm{GC}<200$~kpc.
We identify a triaxial stellar halo (axes ratio $10:8:7$), $43\degree$ tilted from the disk, showing two break radii at $\sim16$~kpc and $\sim76$~kpc, likely associated with Gaia-Sausage/Enceladus (GSE) and Large Magellanic Cloud (LMC), respectively. The {\it inner} stellar halo ($<30$ kpc) is {\it oblate} and aligned with the disk, whereas the {\it outer} stellar halo becomes {\it prolate} and perpendicular to the disk, consistent with the Vast Polar Structure of MW satellites. 
The twisted halo may arise from the disk-halo angular momentum shift triggered by the infall of a massive satellite. 
The anisotropic density distribution of the stellar halo is also measured, with successful re-identification of the Hercules-Aquila Cloud South/North (HAC-N/-S) and Virgo overdensities (VOD). Break radii are found at 15/30~kpc for VOD/HAC-N(-S). We identify the LMC transient density wake with a break radius at 60~kpc in the Pisces overdensity region. We also find new observational evidence of the LMC collective density wake, by showing a break radius at $\sim$100~kpc in the northern Galactic cap with a clear density peak at 90~kpc. In the end, we found that more metal-poor halo stars are more radially extended.
Our results provide important clues to the assembly and evolution of the MW stellar halo under the standard cosmic structure formation framework.
\end{abstract}

\keywords{Galaxy: halo / Galaxy: stellar content / Galaxy: structure}

\section{Introduction}
\label{sec:intro}

The stellar halo of the MW, although comprising only $\sim$2\% of the total stellar mass of the MW \citep{2019MNRAS.490.3426D}, encodes a unique record of its accretion history. According to the $\Lambda$ cold dark matter ($\Lambda$CDM) model, the MW stellar halo is formed hierarchically, and the assembly history is embedded in the accretion debris and in the form of stellar overdensities and streams \citep{1962ApJ...136..748E,1978ApJ...225..357S,1991ApJ...379...52W,1999Natur.402...53H,2002ARA&A..40..487F,2005ApJ...635..931B,2013NewAR..57..100B,2016ARA&A..54..529B,2016ASSL..420..141J,2020ARA&A..58..205H}, including, for example, the Hercules-Aquila Cloud South/North overdensity \citep[HAC-N/-S;][]{2007ApJ...657L..89B,2009MNRAS.398.1757W,2014MNRAS.440..161S}, the Virgo overdensity \citep[VOD;][]{2001ApJ...554L..33V,2002ApJ...569..245N,2008ApJ...673..864J,2023ApJ...951...26C}, and the Pisces overdensity \citep{2007AJ....134.2236S,2009MNRAS.398.1757W,2009ApJ...705L.158K,2015ApJ...810..153N,2023ApJ...951...26C,2023ApJ...956..110C,2024arXiv240817250V}.

Contemporary wide-field surveys, from {\it Gaia} space mission \citep{2016A&A...595A...1G} to large spectroscopic surveys such as SEGUE \citep{2009AJ....137.4377Y}, APOGEE \citep{2017AJ....154...94M}, GALAH \citep{2017MNRAS.465.3203M}, H3 \citep{2019ApJ...883..107C}, LAMOST \citep{2012RAA....12.1197C}, and DESI \citep{desiInstrument,desiScience,desi-collaboration22a,Spectro.Pipeline.Guy.2023,SurveyOps.Schlafly.2023,Corrector.Miller.2023,FiberSystem.Poppett.2024,DESI2023b.KP1.EDR,DESI2024.VII.KP7B,2025arXiv250314738D}, now provide detailed chemical measurements and 6-dimensional phase-space information for millions of stars, and thus the density distribution of the MW stellar halo can be directly quantified by star counting.

Although assembling a whole-sky stellar sample that probes the Galaxy remains challenging, many investigations have charted the density distribution of the MW stellar halo. Various tracers that have precise distance estimates and are luminous enough can be used to measure the density distribution, including main-sequence turn-off \citep[MSTO;][]{2000AJ....119.2254M,2008ApJ...680..295B,2011ApJ...731....4S,2015A&A...579A..38P,70kpc_halo}, blue horizontal branch \citep[BHB;][]{1987MNRAS.227P..21S,1991ApJ...375..121P,2010ApJ...720L.108G,2011MNRAS.416.2903D,2018PASJ...70...69F,2018MNRAS.481.5223T,2019PASJ...71...72F,2019MNRAS.490.5757S,2024ApJ...975...81Y,joao}, RR-Lyrae \citep[RRL;][]{1984MNRAS.206..433H,1991ApJ...375..121P,1996AJ....112.1046W,2006AJ....132..714V,2009MNRAS.398.1757W,2010ApJ...708..717S,2013AJ....146...21S,2014ApJ...788..105F,2018ApJ...859...31H,2018ApJ...855...43M,2019MNRAS.482.3868I,2020MNRAS.497.1547P,2021ApJ...911..109S,2024MNRAS.531.4762M,2025arXiv250402924M}, and red giant branch \citep[RGB;][]{2015ApJ...809..144X,2018MNRAS.473.1244X,2022AJ....164..241Y,Han_stellar_halo_density_profile,2023MNRAS.526.1209L,2024A&A...684A.135L} stars. 
Most studies model the stellar halo as an oblate spheroid aligned with the Galactic disk, fitting its density with a broken power-law profile \citep{2011MNRAS.416.2903D,2022AJ....164..241Y,joao,70kpc_halo}. 
In particular, \cite{2013ApJ...763..113D} revealed that break radii record apocentric pile-ups from early major mergers. \cite{2019ApJ...884...51G} and \cite{2021ApJ...923...92N} further pointed out that major mergers will also induce stellar overdensities in the sky, leading to a highly anisotropic stellar halo.

Recently, multiple works \citep{Han_stellar_halo_density_profile,2022AJ....164..241Y,2023MNRAS.526.1209L} have independently recovered a double-break power-law density profile for the MW stellar halo, with the two break radii corresponding to two apocenter passages of GSE. These discoveries confirmed the dominant role of the GSE merger in shaping the inner $\sim$30~kpc of the MW stellar halo \citep{2018MNRAS.478..611B,2018ApJ...863..113H,2018Natur.563...85H,2019MNRAS.486..378L,2021ApJ...923...92N,2022ApJ...924...23W,2023NatAs...7.1481H,2025ApJ...985L..22N,2025arXiv250402924M}. However, the constraints on the exact locations of break radii and the morphology of MW stellar halo are still subject to large inconsistencies.
Based on the LAMOST survey, \cite{2022AJ....164..241Y} reported an oblate stellar halo without tilt, and the flattening of the stellar halo is changing with radius. However, \cite{Han_stellar_halo_density_profile} (H3) and \cite{2023MNRAS.526.1209L} (APOGEE) have found a prolate stellar halo with tilt. Due to the limited survey depth and sample size, the measurements are mostly within $\sim$50~kpc in these studies.

More recently, and benefited from the usage of large photometric samples to identify BHB fractions, \cite{joao} studied the anisotropic density distributions of the MW stellar halo by fitting the density distribution at different line-of-sight directions.
They also reported that in the outer stellar halo, $\sim$55~\% of stars reside in $\sim$20~\% of sky patches. 
\cite{70kpc_halo} further measured the density profile in these dense sky patches. They detected a clear overdensity in the outer stellar halo, with a break radius at $\sim$70~kpc. However, both \cite{joao} and \cite{70kpc_halo} use photometric samples, which might suffer from more contaminations and imprecise distance measurements than spectroscopic samples \citep{Han_stellar_halo_density_profile,2022AJ....164..241Y,2023MNRAS.526.1209L}.

The Milky Way Survey (MWS) of DESI will release $\sim$12 million stellar spectra in its second data release (DR2). It provides one of the largest and deepest spectroscopic samples of the MW stellar halo spanning a wide footprint, and hence it offers an invaluable spectroscopic data set to investigate the anisotropic density distribution of the MW stellar halo down to unprecedented details.
In this paper, we measure the shape, orientation, and density profile of the MW stellar halo out to $\sim$200~kpc, using the large halo K giant sample from DESI DR2, which is currently an internal version of data release. In particular, we explore how the shape and orientation of the MW stellar halo change with distances, and its alignment with the MW disk and the Vast Polar Structures of MW satellite galaxies, which provides important probes into the assembly and evolution of the MW stellar halo under the standard hierarchical cosmic structure formation scenario. We pay particular attention to the anisotropic density profiles along different directions, and look at the imprints from the perturbations of the Large Magellanic Cloud (LMC) and investigate the dependence of the stellar halo density profiles on [Fe/H].

This paper is organized as follows: Section~\ref{sec:data} gives an overview of the MWS data from DESI and introduces the data selections. We introduce the methodology adopted in this paper in Section~\ref{sec:methodology}, including the selection function of DESI and forward modelling. Section~\ref{sec:results} presents the global best-fit model of the MW stellar halo, the shape and orientation dependence on distances and alignments with the disk and distant satellites, the anisotropic density distribution induced by GSE and LMC, and the model dependence on [Fe/H]. We discuss and compare our results with previous studies in Section~\ref{sec:discussions}, and conclude in Section~\ref{sec:concl}.

\section{Data}
\label{sec:data}
\subsection{The DESI Milky Way Survey}
\label{sec:DESIMWS}

In this paper, we use K giants observed in the \texttt{MAIN-BRIGHT} program of DESI MWS DR2 observation, which is the current internal data release, and will be released in 2027. 
The \texttt{MAIN-BRIGHT} program is one of the main target categories of DESI MWS, which is observed in bright time together with DESI Bright Galaxies \citep{2023AJ....165..253H} and covers a magnitude range of $16<r<19$.

For the \texttt{MAIN-BRIGHT} program of DESI, the targets are further split into a few different classes, including \texttt{MAIN-BLUE} and \texttt{MAIN-RED}. 
\texttt{MAIN-BLUE} subsample contains blue stars ($g-r<0.7$), which mainly probe metal-poor thick disk/halo turn-off stars. \texttt{MAIN-RED} subsample contains red stars ($g-r\geq0.7$), selected with additional astrometric criteria to prioritize the selection and observation of distant halo giants. \texttt{MAIN-BLUE} shares the same observation priority with \texttt{MAIN-RED}. In this work, we only use \texttt{MAIN-BLUE} and \texttt{MAIN-RED} subsamples. We refer readers \cite{2023ApJ...947...37C} for more details of DESI MWS target categories and criteria.

\subsection{MWS stellar parameter catalog}
\label{sec:rvj}

In this work, we use the stellar parameter and radial velocity value-added catalog (VAC) produced by the internal \textsc{rvj} pipeline, which is an improvement on the DESI MWS RVSpecFit (\textsc{rvs}) pipeline used to produce the early data release (EDR) and the first data release (DR1) VACs \citep{2024MNRAS.533.1012K,koposov2025desidatarelease1}. The biggest difference between the \textsc{rvj} and \textsc{rvs} pipelines is that the \textsc{rvj} pipeline uses templates that are synthesized with Korg \citep{2023AJ....165...68W} and provides measurements for individual elemental abundances, including [Mg/Fe], rather than just bulk alpha abundances.

Distances of individual stars are estimated from the DESI \texttt{SpecDis2} VAC (Li et al., in prep.), which is an updated version of DESI DR1 \texttt{SpecDis} VAC \citep{li2025specdisvalueaddeddistance}. The main difference between the \texttt{SpecDis2} and \texttt{SpecDis} catalogs is that \texttt{SpecDis2} predicts distances from stellar spectra and stellar colors, while \texttt{SpecDis} predicts distance modulus only from stellar spectra. Moreover, \texttt{SpecDis2} trains main-sequence and giant stars separately. \texttt{SpecDis2} has significantly increased the precision of distance measurements for giants. The median distance uncertainty of giants decreases from $~$40\% (\texttt{SpecDis}) to $~$12\% (\texttt{SpecDis2}).

\subsection{Selection of K giants}
\label{sec:Selection of K giants}
\begin{figure}
\begin{center}
\includegraphics[width=0.49\textwidth]{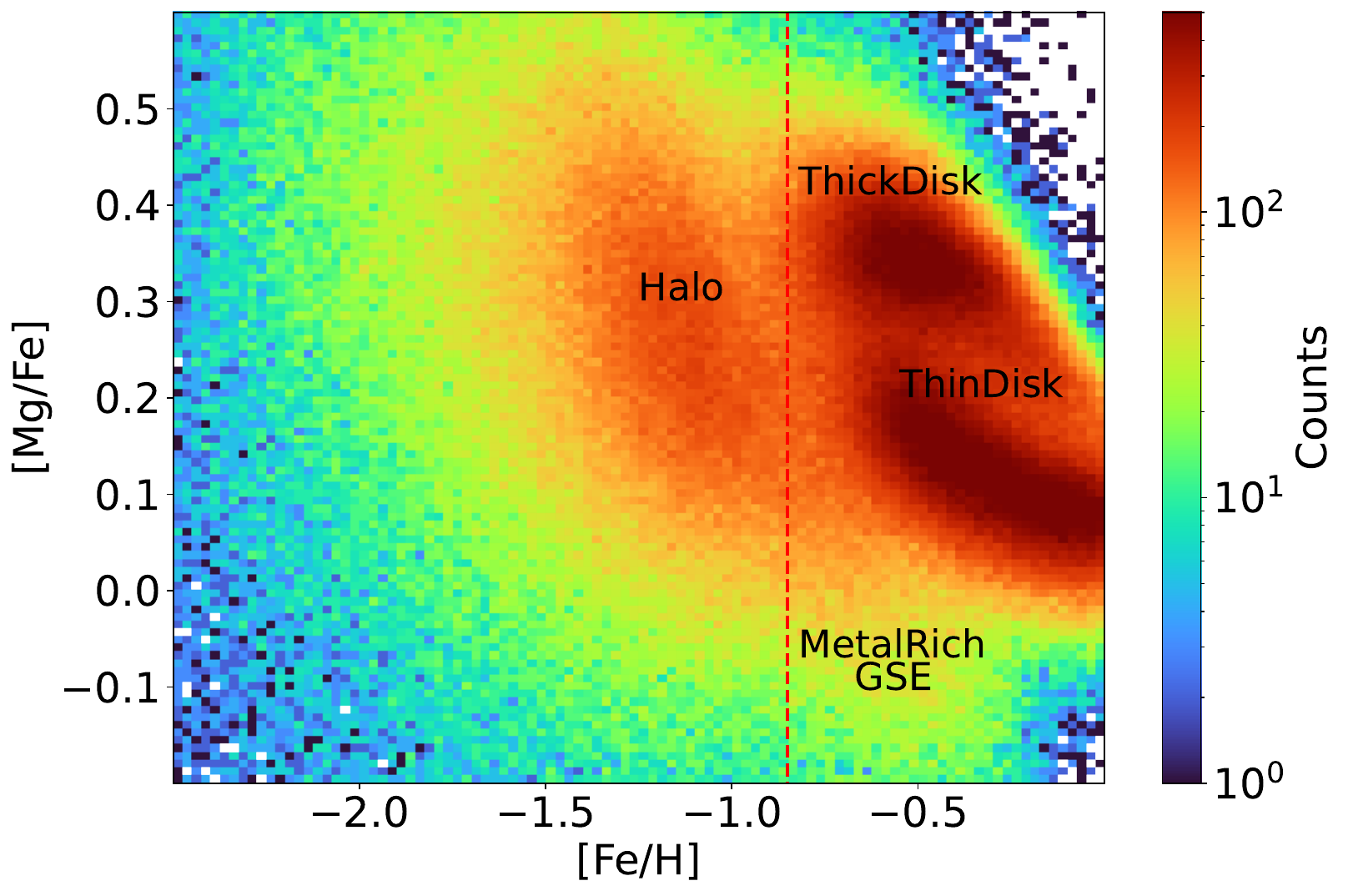}%
\end{center}
\caption{DESI DR2 giants in [Fe/H]-[Mg/Fe] space from the \textsc{rvj} pipeline. The giants are selected from the \texttt{MAIN-BRIGHT} program.
Halo stars are defined by [Fe/H] < -0.85 (vertical red dashed line).}
\label{fig:halo selection}
\end{figure}

In this section, we introduce our K giant sample selection criteria. Our goal is to build a highly pure halo K giant sample to allow robust measurements of the shape, orientation, and radial density profile of the stellar halo. 

Based on the DESI MWS \texttt{MAIN-BLUE} and \texttt{MAIN-RED} subsamples, we first select K-type stars with colors in the range of $0.5<g-r<1.3$ after extinction correction. With this selection, we get a sample that contains $\sim$135,000 stars. 
We further select K giants by absolute magnitude \citep{huang2023spectroscopyiistellarparameters}. Our K giants are defined by $-3<M_{r}<2$ ($r$-band absolute magnitude after extinction correction), which removes various types of faint stars. After this selection, the remaining sample contains $\sim$92,000 stars\footnote{We note that \texttt{MAIN-RED} targets are incomplete when close to the Sun, because stars with $\omega>3\sigma_{\omega}+0.3\mathrm{mas}$ are not included as \texttt{MAIN-RED} targets. However, all K giants in our final sample are beyond 6~kpc from the Sun due to our chosen absolute magnitude range above. The incompleteness of \texttt{MAIN-RED} targets would not affect our results.}. 

Most of these K giants selected through the steps above lie at $>1~\mathrm{kpc}$ above the disk plane, which is due to the flux limit of the survey ($16<r<19$) and magnitude cut we adopt here, which has already removed the majority of contamination from the disk. We further remove stars with $\mathrm{[Fe/H]}>-0.85\,\mathrm{dex}$ (see red dashed line in Figure~\ref{fig:halo selection}). This may also remove a small number of metal-rich GSE stars (see the right-bottom corner of Figure~\ref{fig:halo selection}), but ensures a purer halo star sample \citep{2015ApJ...808..132H}. After this selection, we retain about 46,000 stars.

As the final requirement in our selection, we remove stars belonging to known substructures. Member stars from globular clusters (GCs) and dwarf galaxies are removed by matching specific stars with \cite{2021MNRAS.505.5957B} and \cite{2022ApJ...940..136P}.
The contribution from the Sagittarius (Sgr) stream is removed by footprint. We compute the latitude coordinate, $\tilde{B}$, perpendicular to the plane of the Sagittarius stream according to the coordinate system defined by \cite{10.1093/mnras/stt1862} for all stars in the sample, and remove stars with $|\tilde{B}|<15$. 
After this, we get a final halo K giant sample which contains $\sim$28,000 stars, which covers a galactocentric distance ($r_\mathrm{GC}$) from 8~kpc to 200~kpc.

\begin{figure*}
\begin{center}
\includegraphics[width=0.78\textwidth]{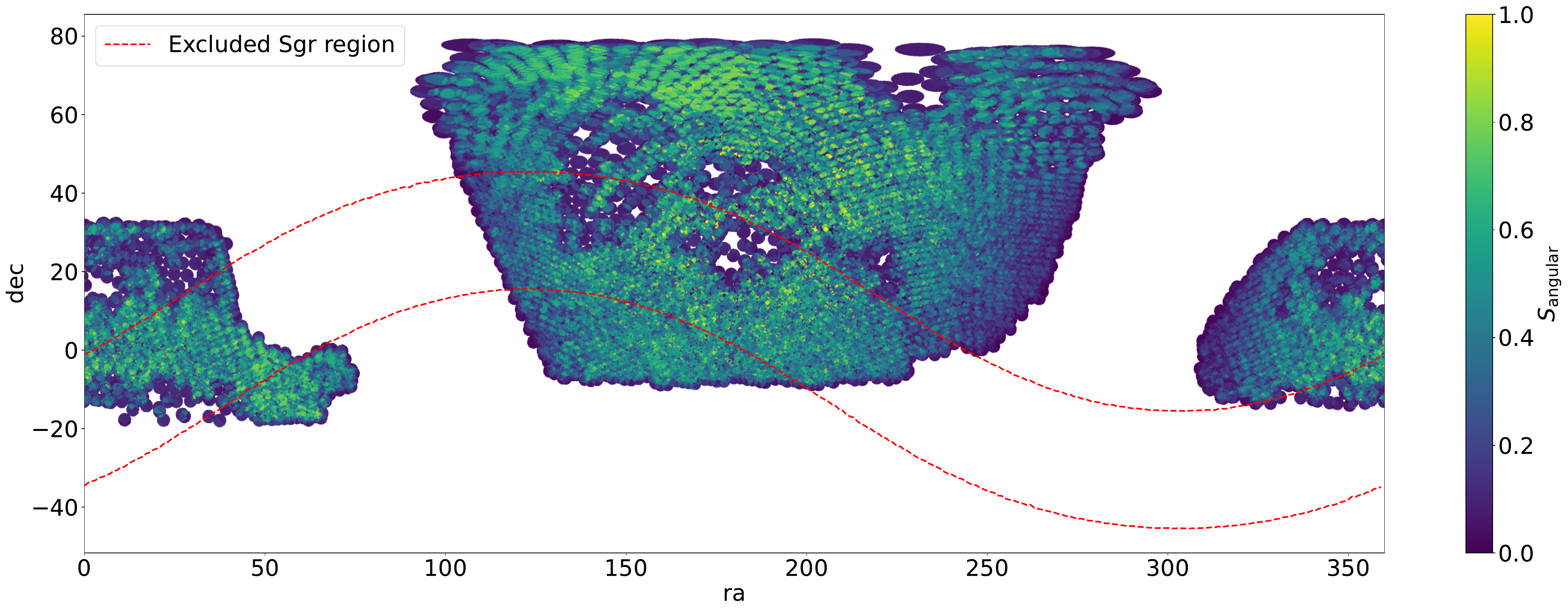}%
\end{center}
\caption{Angular selection function of \texttt{MAIN-BLUE} subsample as a function of sky position. The sky region between the two red dashed lines denotes the Sgr footprint, which has been cut off from our analysis. Here, the angular selection function is defined as the completeness fraction of DESI observed stars with respect to the targets. Brighter colors indicate higher completeness. 
\texttt{MAIN-RED} subsample shares a similar angular selection function with \texttt{MAIN-BLUE}. We do not repeatedly present the angular selection function of \texttt{MAIN-RED} subsample.}
\label{fig:selection of blue}
\end{figure*}

\section{Methods}
\label{sec:methodology}
We take two major steps to construct the spatial distribution of the stellar halo with K giants: 1) correcting the selection effects due to incompleteness of the spectroscopic survey and incompleteness of photometric targets due to the flux limit, and 2) reconstructing the stellar halo with forward modelling, which contains a set of flexible and smooth functions with free parameters.

\subsection{Correction of Incompleteness}
\label{sec:selection function}
Our K giant sample is incomplete for two reasons. The first is incompleteness in spectroscopically observed stars with respect to photometric targets. The second is a consequence of the MWS target selection, as the \texttt{MAIN-BRIGHT} program only observed stars with $16<r<19$, and thus K giants with different ranges of absolute magnitudes are selected at different distances, resulting in an incompleteness that is radially dependent. We correct these two selection effects with two selection functions.

\subsubsection{Correcting Incompleteness of Spectroscopic Survey}
\label{sec:Correcting Incompleteness of Spectroscopic Survey}
The first selection function is defined as the completeness fraction of spectroscopically observed stars with respect to the photometric targets, which can be calculated by comparing the spectroscopically observed stars with the photometric input targets. In our final halo K giant sample, stars with $0.5<g-r<0.7$ and $0.7<g-r<1.3$ belong to \texttt{MAIN-BLUE} and \texttt{MAIN-RED}, respectively. 
We then construct two selection functions for \texttt{MAIN-BLUE} and \texttt{MAIN-RED} separately\footnote{It is important to treat \texttt{MAIN-BLUE} and \texttt{MAIN-RED} separately, as they correspond to different photometric targets.}.
According to \cite{2017RAA....17...96L}, the selection function ($S$) usually depends on the sky position, color, and apparent magnitude, while
we further check that for both \texttt{MAIN-BLUE} and \texttt{MAIN-RED}, the selection function in fact shows almost no dependence on color or apparent magnitude. Thus, for both \texttt{MAIN-BLUE} and \texttt{MAIN-RED} sources, the selection functions reduce to being only dependent on sky coordinates:
\begin{equation}
S(\mathrm{obs}|\mathrm{ra},\mathrm{dec},\mathrm{class})=\frac{n_\mathrm{sp,\mathrm{class}}(\mathrm{ra},\mathrm{dec})}{n_\mathrm{ph,\mathrm{class}}(\mathrm{ra},\mathrm{dec})}.
\label{equ:selection function1}
\end{equation}

Here ``class'' refers to \texttt{MAIN-BLUE} or \texttt{MAIN-RED}, ``obs'' stands for observed by DESI, and ra, dec are two celestial coordinates. $n_\mathrm{sp}$ and $n_\mathrm{ph}$ are the number of spectroscopically observed stars by DESI and the photometric targets, as a function of sky coordinates. We call it the angular selection function.

To calculate the angular selection function, we first divide the whole footprint of DESI DR2 MWS data into different regions, and each region has a different number of passes, called $N_\mathrm{passes}$. This parameter indicates the number of DESI tiles at a given sky position\footnote{For DESI bright time observation, the tiles are organized into five passes. Each tile corresponds to a single arrangement of DESI fibers to observe sources. Tiles within a pass do not overlap, whereas tiles on different passes are offset such that all points on the sky will be covered by three tiles on average \citep{2023ApJ...947...37C}, i.e., the mean $N_\mathrm{passes}$ is three. However, most MWS sources are observed only once, as bright galaxies are also observed at bright time and are assigned higher fiber priorities than most MWS sources.}. Naturally, $N_\mathrm{passes}$ is a function of ra and dec, and a higher $N_\mathrm{passes}$ will lead to a higher completeness fraction and thus a higher selection function value. For regions with the same $N_\mathrm{passes}$, we pixelize them with the Python package \texttt{healpy} \citep{healpy}, and count the number of spectroscopic and photometric stars in each pixel to get the selection function in different pixels.
As an example, we present the selection function of the \texttt{MAIN-BLUE} subsample. Figure~\ref{fig:selection of blue} shows the angular selection function of \texttt{MAIN-BLUE} as a function of sky position. The tiling patterns are well captured.

\subsubsection{Correcting Incompleteness of Photometric Targets}
\label{sec:Correcting Incompleteness of Photometric Targets}
In this section, we introduce the second selection function that corrects the incompleteness of photometric targets due to the survey flux limit. Since the flux limit affects the distance range that we can observe for a source with fixed absolute magnitude, we call it the radial selection function.

We assume that the luminosity function of K giants obeys an exponential law $\mathrm{\Phi\,(M_\mathrm{r})\sim10^{0.32 M_r}}$ \citep{2014ApJ...784..170X}. For a given distance $D(\mathrm{kpc})$ to the Sun, the apparent magnitude of K giants covers a range from $r=7+5\,\log D$ (corresponding to an absolute magnitude of $M_r=-3$) to $r=12+5\,\log D$ (corresponding to an absolute magnitude of $M_r=2$). The fraction of a K giant that has been targeted is: 
\begin{flalign}
\label{equ:selection function2}
&&
\begin{aligned}[t]
S_{\mathrm{radial}}(\mathrm{obs}|D,m_1,m_2,\mathrm{class})=&\\
{\frac{\int_{m_1}^{m_2} \Phi(r-5\log D-10)\,\mathrm{d}r}
       {\int_{7+5\log D}^{12+5\log D} \Phi(r-5\log D-10)\,\mathrm{d}r}}&
\end{aligned}
&&
\end{flalign}
where $m_1$ is the larger one of 16 and $7+5\,\log D $, and $m_2$ is the smaller of 19 and $12+5\,\log D$. Again, class refers to \texttt{MAIN-BLUE} or \texttt{MAIN-RED}. The typical value of $S_{\mathrm{radial}}$ is about 0.4 at $D\sim30~\mathrm{kpc}$. $S_{\mathrm{radial}}$ decreases quickly with distances, which becomes $\sim0.03$ at $D\sim50~\mathrm{kpc}$.

With the angular and radial selection functions, we can correct for the incompleteness in our sample of K giants. This is achieved by assigning each star a weight, which is the product of the inverse of the angular and radial selection functions.

\subsection{Constructing the Stellar Halo with Forward Modelling}
\label{sec:parameterized_model}

In this section, we introduce our parameterized forward model, which describes the shape, orientation, and radial density profile of the MW stellar halo. 

We assume that the stellar halo is a triaxial ellipsoid. The shape of the ellipsoid can be fully described by the flattened radius:
\begin{equation}
\label{equ:flattened_radius}
r_q \equiv \sqrt{X^2 + \left(\frac{Y}{p}\right)^2 + \left(\frac{Z}{q}\right)^2}, 
\end{equation}
where $X$, $Y$, and $Z$ are Cartesian coordinates centered on the galactic center, and $p$, $q$ are intermediate axis flattening and minor axis flattening ($0<q<p<1$), respectively. Here, we allow the $X-, Y-,$ and $Z-$axes to be rotated with respect to the standard galactocentric Cartesian coordinates ($x$,$y$,$z$), with $z$ perpendicular to the Galactic disk, and $x$ points from the Sun toward the Galactic center. Hence, in principle, we need three rotation angles to fix the orientation of the ellipsoid in our modelling: the rotation angle around the minor axis ($\phi$, yaw angle), the rotation angle around the intermediate axis ($\theta$, pitch angle), and the rotation angle around the major axis ($\zeta$, roll angle). In this work, we adopt the convention that a positive value of the rotation angle indicates clockwise rotation.

Importantly, the pitch angle $\theta$ is the angle between the major axis of the stellar halo and the Galactic plane. The yaw angle $\phi$ is the angle between the vector connecting the Sun to the Galactic center and the projection of the model major axis on the Galactic plane. 
However, \cite{Han_stellar_halo_density_profile} found a prolate-like stellar halo, and hence failed to get a good fitting of the roll angle. In this work, we also find that the stellar halo is nearly prolate with a similar axis ratio to that in \cite{Han_stellar_halo_density_profile} (see Section~\ref{sec:results}). We also try to fit the roll angle, and the best-fit value is nearly zero. Hence, in this work, we choose to only fit the pitch angle and yaw angle of the stellar halo, by fixing the roll angle to zero throughout our analysis.

Following \cite{Han_stellar_halo_density_profile}, we assume that the radial density profile of the MW stellar halo obeys a triple power-law with two break radii
\begin{equation}
\label{equ: triple power law}
\rho\,(r_q) \propto 
\begin{cases} 
r_q^{-\alpha_1} & \text{; } r_q \leq r_{b,1}, \\
r_{b,1}^{\alpha_2 - \alpha_1} r_q^{-\alpha_2} & \text{; } r_{b,1} < r_q \leq r_{b,2}, \\
r_{b,2}^{\alpha_3 - \alpha_2} r_q^{-\alpha_3} & \text{; } r_{b,2} < r_q,
\end{cases}
\end{equation}
where $\alpha_1$, $\alpha_2$, $\alpha_3$ are the first, second, and third power-law slopes, and $r_{b,1}$ and $r_{b,2}$ are the first and second break radii. We also test a less flexible radial density profile with only one break radius and two power-law slopes:
\begin{equation}
\label{equ: double power law}
\rho\,(r_q) \propto 
\begin{cases} 
r_q^{-\alpha_1} & \text{; } r_q \leq r_{b,1}, \\
r_{b,1}^{\alpha_2 - \alpha_1} r_q^{-\alpha_2} & \text{; } r_{b,1} < r_q , \\
\end{cases}
\end{equation}
to compare with the triple power-law model. In total, there are nine free parameters for the triple power-law model (two in shape, two in orientation, and five in density profile) and seven for the second power-law model, which lacks two density profile parameters. We have also tried other, more asymptotically changing functional forms, but have failed to see significant changes in the best-recovered positions of the break radii.

In addition to the above triple power-law model, we also adopt a model that allows the shape and orientation to change with distances, which is called the variable model\footnote{For the variable model, we fix the radial density profile to be a triple power-law model, with parameters from those of the best-fit model obtained in Section~\ref{sec:triple power-law model results}}. We assume that the $p$, $q$, $\theta$, and $\phi$ parameters change with galactocentric radius, $r_\mathrm{GC}$, as 3rd-order polynomials: 
\begin{equation}
\mathrm{p}\,(r_\mathrm{GC}) = \mathrm{p}_0+k_{\mathrm{p}1}r_\mathrm{GC}+k_{\mathrm{p}2}r_\mathrm{GC}^2+k_{\mathrm{p}3}r_\mathrm{GC}^3,
\label{equ:variable_model}
\end{equation}
\begin{equation}
\mathrm{q}\,(r_\mathrm{GC}) = \mathrm{q}_0+k_{\mathrm{q}1}r_\mathrm{GC}+k_{\mathrm{q}2}r_\mathrm{GC}^2+k_{\mathrm{q}3}r_\mathrm{GC}^3,
\end{equation}
\begin{equation}
\phi\,(r_\mathrm{GC}) = \phi_0+k_{\phi1}r_\mathrm{GC}+k_{\phi2}r_\mathrm{GC}^2+k_{\phi3}r_\mathrm{GC}^3,
\end{equation}
\begin{equation}
\theta\,(r_\mathrm{GC}) = \theta_0+k_{\theta1}r_\mathrm{GC}+k_{\theta2}r_\mathrm{GC}^2+k_{\theta3}r_\mathrm{GC}^3.
\end{equation}
$\mathrm{p}_0$, $\mathrm{q}_0$, $\phi_0$, and $\theta_0$ are axes flattening parameters and rotation angles at $r_\mathrm{GC}=0$. $k_{\mathrm{p}i}$, $k_{\mathrm{q}i}$, $k_{\phi i}$, and $k_{\theta i}$ are polynomial coefficients, and i is in the range 1 to 3. There are 16 free parameters in the variable model.

We will present the best-fit triple power-law model in Section~\ref{sec:triple power-law model results} and compare it with the best-fit double power-law model. We further show the best-fit variable model in Section~\ref{sec:variable results}.

\begin{figure*}
\begin{center}
\includegraphics[width=0.8\textwidth]{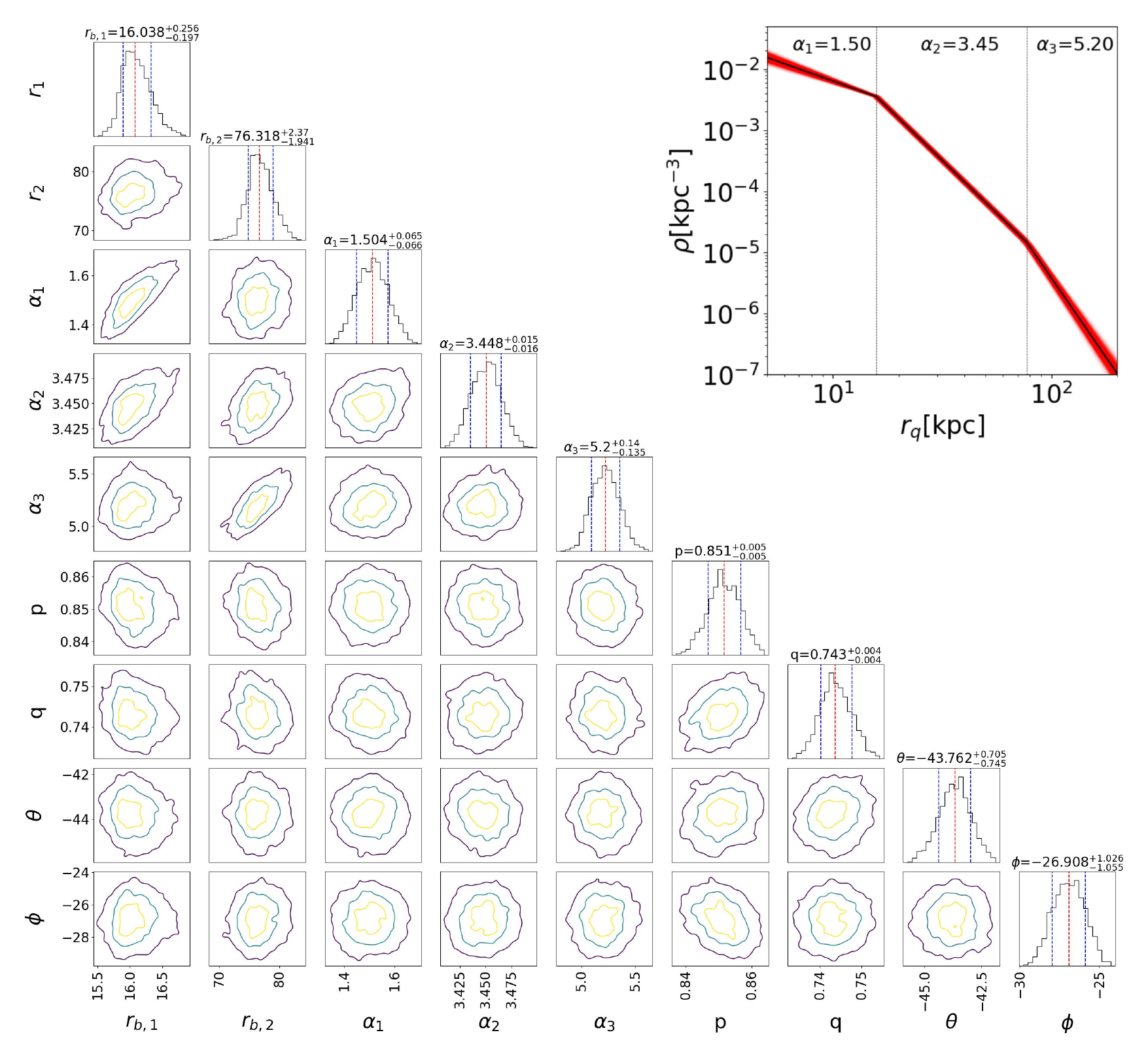}%
\end{center}
\caption{Posterior contours for different combinations of nine parameters of the triple power-law model. Red and blue dashed lines indicate 50th, 16th, and 84th percentiles. The meanings of different model parameters can be found in Section~\ref{sec:parameterized_model}. The yellow, light blue, and dark blue contours represent the $30\%$, $1\sigma$, and $2\sigma$ regions of the MCMC post-burn distributions, respectively. The black line in the upper right panel shows the selection effect free best-fit model radial density profile (renormalized to unity) with the red shaded region representing the 1-$\sigma$ model uncertainty, and two vertical black dashed lines mark two break radii.}
\label{fig:mcmc_2rb}
\end{figure*}

\begin{figure*}
\begin{center}
\includegraphics[width=0.98\textwidth]{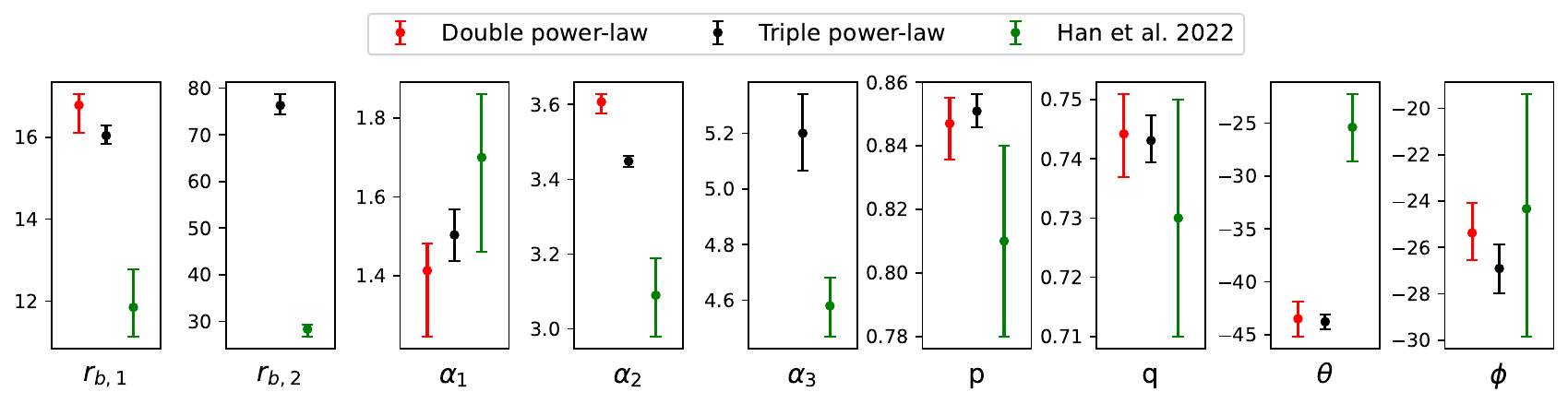}%
\end{center}
\caption{Best-fit model parameters and uncertainties for the double power-law model (red), the triple power-law model in this work (black), and the triple power-law model (green) in \cite{Han_stellar_halo_density_profile}. The errorbars represent the 1-$\sigma$ uncertainties of three models.
The meanings of different model parameters can be found in Section~\ref{sec:parameterized_model}. Three models share similar flattening parameters and yaw angle (the difference is within 1-$\sigma$ uncertainty), while they show a greater discrepancy in radial density profile parameters.}
\label{fig:compare_with_han}
\end{figure*}

\begin{figure}
\begin{center}
\includegraphics[width=0.45\textwidth]{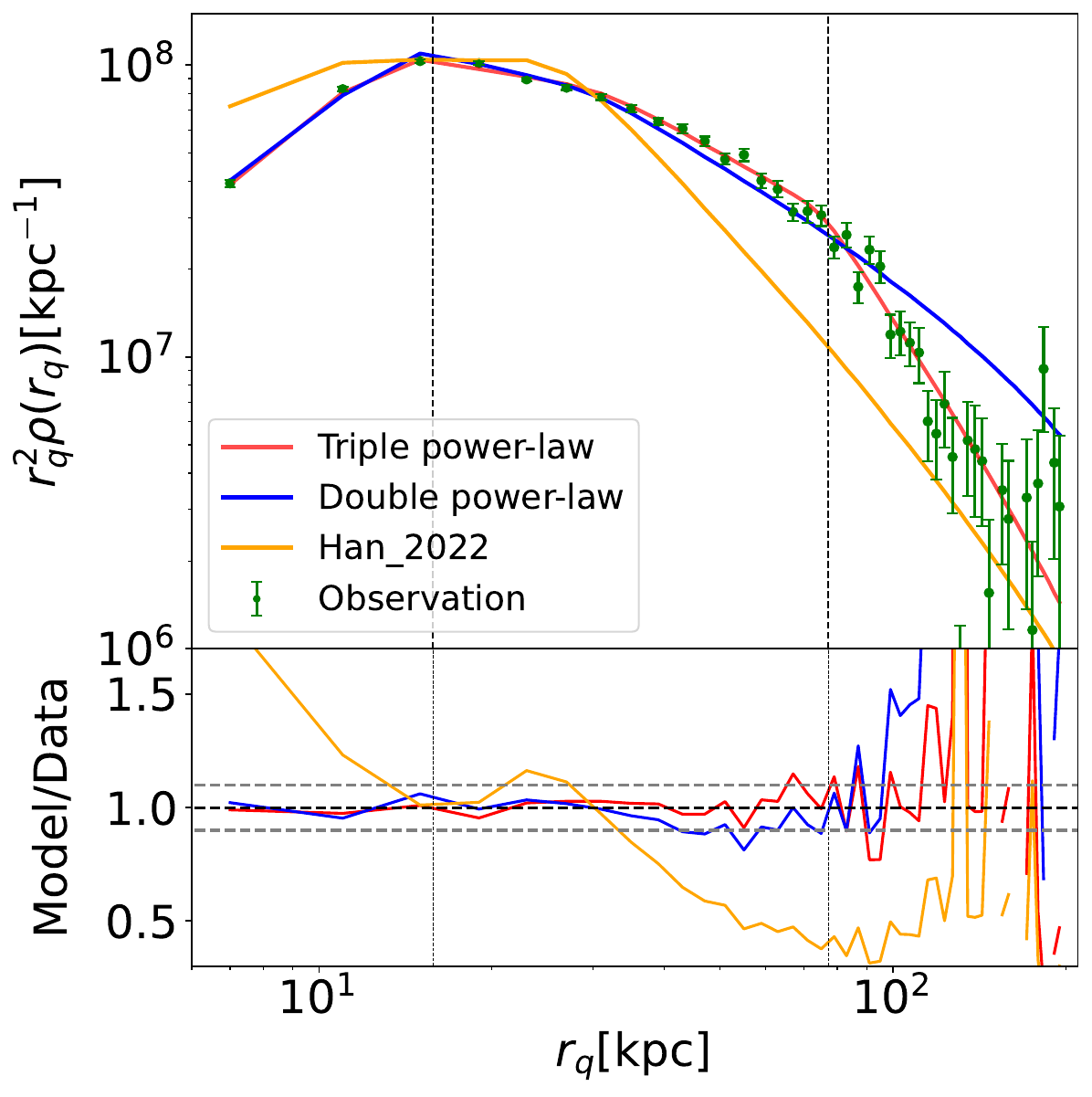}%
\end{center}
\caption{The upper panel shows in green dots the observed stellar density multiplied by $r_q^2$ as a function of flattened radius, $r_{q}$, compared to the best-fit triple power-law model (red solid line) and double power-law model (blue solid line). Errorbars represent the 1-$\sigma$ uncertainties computed from 100 bootstrap subsamples of our K giants. Here, the best-fit triple power-law model in this work (red solid line), that in \cite{Han_stellar_halo_density_profile} (orange solid line), and the double power-law model (blue solid line)
have been convolved with the angular and radial selection functions to have a fair direct comparison with the data (see Section~\ref{sec:selection function}). 
Two vertical black dotted lines in both upper and bottom panels represent the two break radii, $r_{b,1}$ and $r_{b,2}$, respectively. Two horizontal dashed gray lines in the bottom panels represent 10~\% regions of the model-predicted density over observed density, with the black dashed horizontal line marking $y=1$. Both triple power-law and double power-law models fit the stellar halo within 70~kpc well, but the double power-law model shows a worse match of the outer stellar halo beyond $\sim$ 70~$\mathrm{kpc}$.}
\label{fig:1rb_2rb_compare}
\end{figure}

\subsection{Likelihood Function}
\label{likelihood function}

With angular and radial selection functions and a forward model for the radial density profile above, the probability for the $i$-th K giant with a given heliocentric distance, $D_i$, sky coordinate $\mathrm{ra_i}$ and $\mathrm{dec_i}$ to be observed by DESI is:
\begin{equation}
P_i(D_i,\mathrm{ra_i,dec_i}) = \frac{S_{i,\mathrm{angular}}S_{i,\mathrm{radial}}\rho\,(r_{q,i})4\pi r_{q,i}^2}{C} ,
\label{equ:probability}
\end{equation}
\begin{equation}
\label{equ:normalization}
C=\int_\Omega\int_D  S_{\mathrm{angular}}S_{\mathrm{radial}}\rho(r_{q})D^2\mathrm{d}\Omega \mathrm{d}D ,
\end{equation}
where $C$ is the normalization and integrates over the volume in the $(D, \mathrm{ra}, \mathrm{dec})$ space. $S_{i,\mathrm{angular}}$ and $S_{i,\mathrm{radial}}$ refer to the selection functions defined in Equations~\ref{equ:selection function1} and \ref{equ:selection function2}. Equation~\ref{equ:probability} takes the same form for stars in either \texttt{MAIN\_BLUE} or \texttt{MAIN\_RED}, adopting the corresponding \texttt{MAIN\_BLUE} or \texttt{MAIN\_RED} selection functions, respectively. Then we can directly calculate the likelihood of observed stars
\begin{equation}
L = \prod_{i=1}^{N_{\text{K}}} P_i \big(\mathrm{ra}_i,\mathrm{dec}_i,D_{i}),
\label{eq:likelihood}
\end{equation}
where $N_K$ is the total number of K giants and thus the multiplication goes through all K giants in both \texttt{MAIN\_BLUE} and \texttt{MAIN\_RED}. We then assume a uniform prior for all parameters in the stellar halo model and use the Python package \texttt{emcee} \citep{emcee} to sample the posterior distribution. We employ 50 walkers, discard the first 400 steps as before burn-in\footnote{The autocorrelation time is about 35 steps, so the post-burn-in samples can be treated as statistically independent draws from the target distribution \citep{emcee}.}, and run 1000 steps in total.

\section{Results}
\label{sec:results}
In this section, we first present the results of the best-fit triple power-law model for the MW stellar halo (see Section~\ref{sec:triple power-law model results}). We then show the results of the best-fit variable model (see Section~\ref{sec:variable results}). We further investigate the spatially anisotropic stellar halo by measuring the density profiles along different pointings on the sky, including regions associated with the GSE debris and the LMC wake (see Section~\ref{sec:Anisotropic stellar halo density distribution and the LMC wake}). Finally, we explore the model dependence on [Fe/H] by fitting the triple power-law model in various [Fe/H] ranges (see Section~\ref{sec:Model dependence on Metallicity}).
We have verified in Appendix~\ref{app:error_on_model} that distance uncertainties of K giants do not significantly impact our results.

\subsection{Best-fit Triple Power-law Model}
\label{sec:triple power-law model results}

Figure~\ref{fig:mcmc_2rb} shows the posterior contours of each parameter of the triple power-law model. The best-fit parameters and their associated uncertainties are shown in the first row of Table~\ref{table:regions}. For our best-fit model, we find that: (1) The axes ratios of the stellar halo are about $1:p:q = 10:8:7$. Hence, the stellar halo is nearly prolate with $p$ and $q$ more similar to each other. (2) The stellar halo is tilted by about 44$\degree$ off the Galactic plane and about 27$\degree$ away from the Sun-Galactic center axis. (3) The stellar halo has two break radii at 16~kpc and 76~kpc (see inset panel of Figure~\ref{fig:mcmc_2rb}). 

Figure~\ref{fig:compare_with_han} shows the comparison between our best-fit parameters of the triple power-law model (Equation~\ref{equ: triple power law}), and the best fits reported by \cite{Han_stellar_halo_density_profile}. 
Our shape parameters $p$ and $q$ are similar to \cite{Han_stellar_halo_density_profile}. We also found a similar yaw angle $\phi$, while the pitch angle $\theta$ is more negative than \cite{Han_stellar_halo_density_profile}. The discrepancy in orientation parameters may be explained as: (1) DESI and H3 surveys have different footprints. (2) \cite{Han_stellar_halo_density_profile} removed Sgr by angular momenta\footnote{We have also tested removing Sgr by angular momentum. However, there is still some contamination after the angular momenta selection. Besides, Kizhuprakkat et al. (in prep) utilize a K giants sample from DESI MWS DR2 for studying the assembly history of the stellar halo. They further confirm that distance errors scatter Sgr members into retrograde and low-energy regions, blending them with GSE, leading to less robust removal of Sgr only by angular momenta.}, while we mask the footprint of the Sgr stream, which adds a difference in footprint. 

However, we find a more prominent discrepancy in the radial density profile. 
The first break radius $r_{b,1}$ is larger than that in \cite{Han_stellar_halo_density_profile}, and we do not find a break radius around 30~kpc like \cite{Han_stellar_halo_density_profile}, but instead we find a much greater second break radius $r_{b,2}$ at $\sim$70~kpc. Our radial density profile is also steeper than \cite{Han_stellar_halo_density_profile} beyond the first break radius. 

The discrepancy in the radial density profile is mainly caused by the different data selection compared with \cite{Han_stellar_halo_density_profile}. \cite{Han_stellar_halo_density_profile} mainly focused on the radial density profile of GSE stars, by adopting chemical and kinematic selections to obtain a pure GSE sample. Their sample is shallower ($r_\mathrm{GC}<60~\mathrm{kpc}$), and thus \cite{Han_stellar_halo_density_profile} did not report a break radius at $\sim$70~kpc.
In this work, we use all halo stars. The selection on [Fe/H] (see Section~\ref{sec:Selection of K giants}) further removes some GSE stars, leading to a slightly smaller fraction of GSE in our halo sample. However, if we only fit the radial density profile of HAC-N and HAC-S, which are dominated by GSE stars \citep{2019MNRAS.482..921S,2022ApJ...936L...2P}, we can also find a break radius at $\sim$30~kpc (see Section~\ref{sec:Anisotropic stellar halo density distribution and the LMC wake}).
Interestingly, \cite{70kpc_halo} utilized a K giant and MSTO sample which can extend to about 100~kpc, and also found a break radius at $\sim$70~kpc in the southern sky.

Figure~\ref{fig:compare_with_han} also shows the comparison between the parameters and uncertainties of the best-fit double power-law model (Equation~\ref{equ: double power law}) and triple power-law model. The best-fit double power-law model is also obtained from MCMC, for which we do not show the posterior contours. The double power-law model shows a similar shape, orientation, and radial density profile parameters except for the second power-law slope $\alpha_{2}$. The difference between $\alpha_{2}$ of the two models is greater than the model uncertainties, meaning a third component is required to model the MW stellar halo. We have also employed the Akaike Information Criterion\footnote{$\mathrm{AIC} = 2k - 2\ln\hat{L}$, where k is the number of free parameters and $\ln\hat{L}$ is the maximum likelihood; lower AIC favors a better trade-off between fit quality and model complexity.} \citep[AIC;][]{1974ITAC...19..716A}, and the AIC difference between the triple power-law model and the double-power law model is -216, which means that the data prefers the triple power-law model.

Figure~\ref{fig:1rb_2rb_compare} presents the direct 1-dimensional comparison between the observed and the best-fit double/triple power-law model profiles, reported as a function of the flattened radius, $r_{q}$. As a comparison, we also plot the best-fit profile from \cite{Han_stellar_halo_density_profile}. 
In the upper panel, the green dots present the observed stellar density\footnote{The density is obtained from dividing the number of stars in each ellipsoidal shell by its volume.} profile multiplied by $r_q^2$.
The errorbars give 1-$\sigma$ uncertainties estimated from 100 bootstrap resamplings of our K giants. 
The red and blue lines represent the predicted density profile from the triple and double power-law models, for which we convolve them with the DESI footprint and selection functions. The bottom panel of Figure~\ref{fig:1rb_2rb_compare} presents the ratio between the best-fit model profiles and the real data as a function of $r_{q}$. 

Both the triple and double power-law models represent good fits to the observation within 70~kpc. The differences between the two models and the observed density profile are all smaller than 10~\%, with the double power-law model showing slightly larger bias. However, the double power-law model fails to fit the observed density profile and predicts an overestimated density profile beyond 70~kpc, while the triple power-law model captures the transition of density profile with a second break radius $r_{b,2}$ around 70~kpc. The triple power law better describes the radial density profile of the stellar halo than the double power-law model beyond 70~kpc.

\subsection{Radially Variable Model with Fixed Radial Density Profile}
\label{sec:variable results}
\begin{figure*}
\begin{center}
\includegraphics[width=0.95\textwidth]{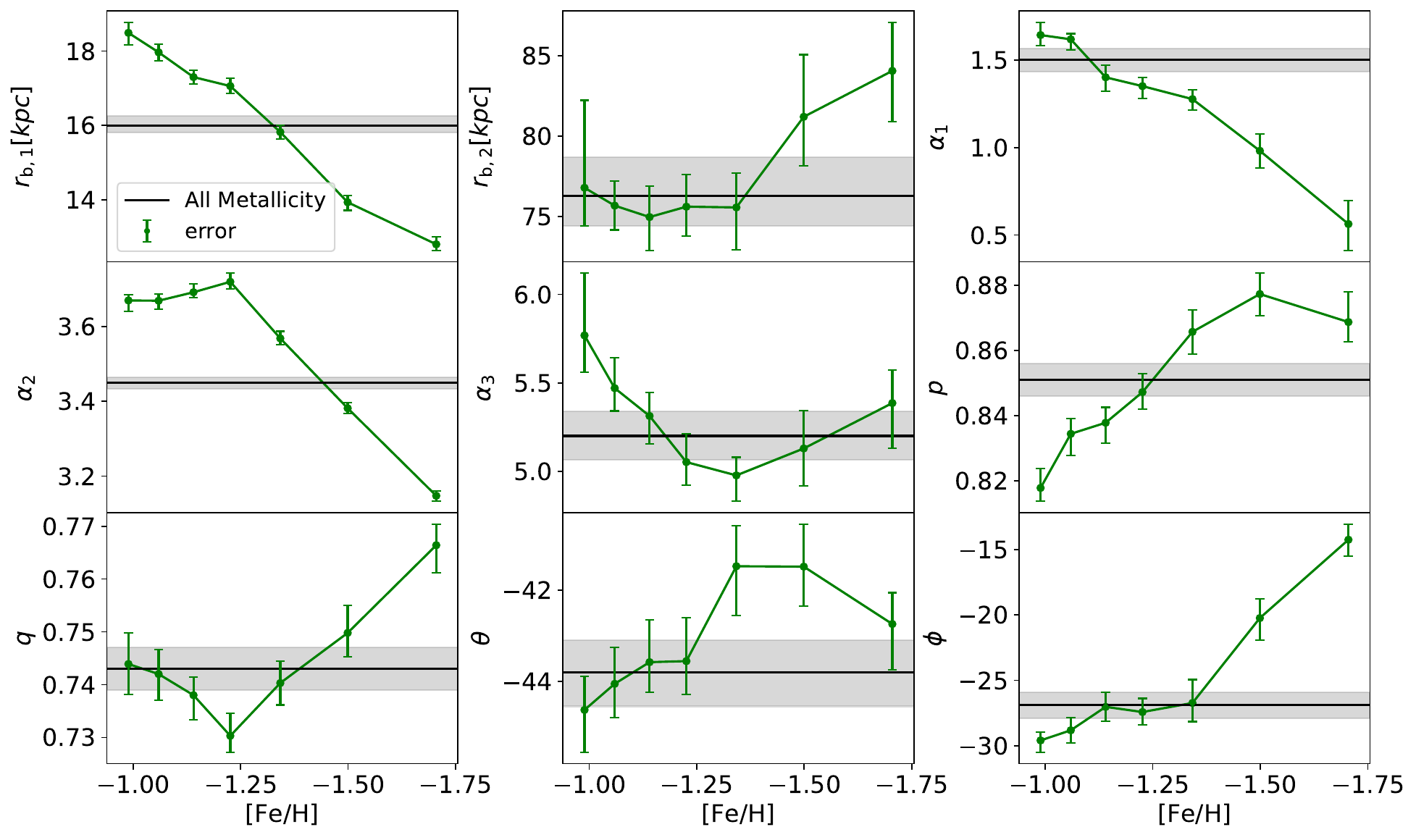}%
\end{center}
\caption{Dependence of the triple power-law model parameters on [Fe/H]. Each green dot presents a subsample of K giants in a given [Fe/H] range, with the errorbars showing the1-$\sigma$ uncertainties of model parameters. The $x$-axis value of each green dot is the median [Fe/H]. The black horizontal solid line is the best-fit model parameter of the triple power-law model in Section~\ref{sec:triple power-law model results}, with the gray shaded region representing the model uncertainties.}
\label{fig:parameters_vs_feh}
\end{figure*}

In this section, we explore a more flexible radially variable stellar halo model, which allows the shape and orientation parameters to change with the radius (see details in Section~\ref{sec:parameterized_model}). We only explore this model with the whole sample, as Figure~\ref{fig:parameters_vs_feh} above has shown that the dependence of shape and orientation parameters on [Fe/H] is relatively weak.

\begin{figure*}
\begin{center}
\includegraphics[width=0.95\textwidth]{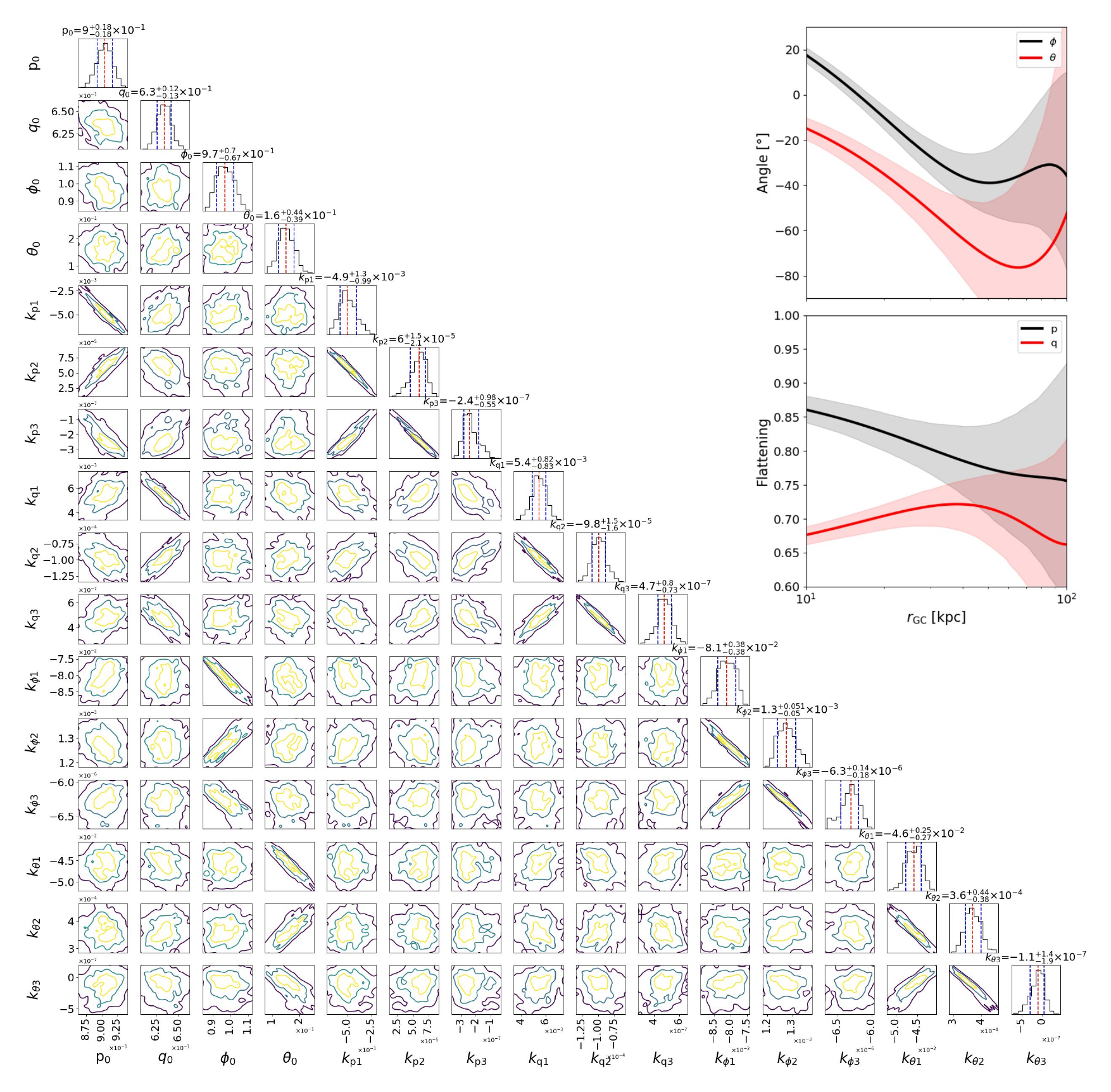}%
\end{center}
\caption{The posterior contours of sixteen model parameters of the variable model. Red dashed lines and blue dashed lines indicate 50th, 16th, and 84th percentiles. The meanings of different model parameters can be found in Sections~\ref{sec:parameterized_model}. The yellow, light blue, and dark blue contours represent the $30\%$, $1\sigma$, and $2\sigma$ regions of the MCMC post-burn distributions, respectively. The two upper right panels present the dependence of the rotation angles $\phi$, $\theta$, and flattenings $p$, $q$ on $r_\mathrm{GC}$. The thick lines are the predictions from the best-fit model, and the shaded regions present the 1-$\sigma$ uncertainty of the variable model.}
\label{fig:mcmc_variable}
\end{figure*}

\begin{figure*}
\begin{center}
\includegraphics[width=0.95\textwidth]{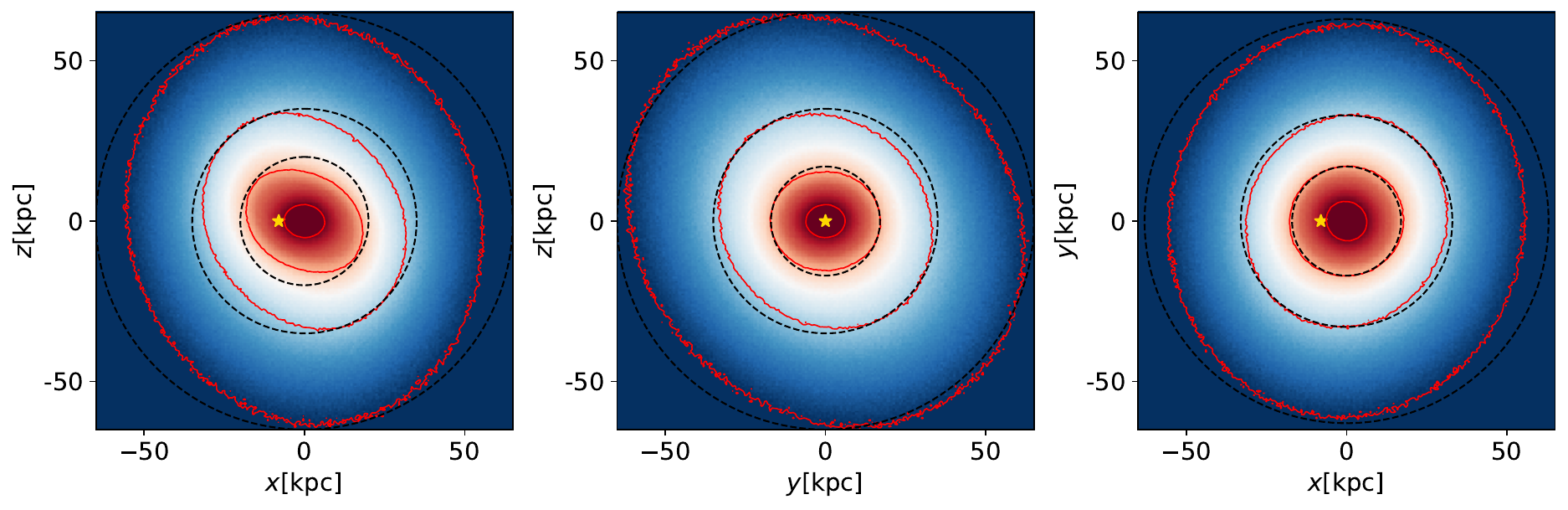}%
\end{center}
\caption{The density distribution of the best-fit variable model in the $x$-$z$ ($y=0$), $y$-$z$ ($x=0$), and $x$-$y$ ($z=0$) slices. Here $x,y,z$ are standard Galactocentric Cartesian coordinates. The red solid lines are isodensity contours, with the density for each outer contour decreased by a factor of 10 compared with the previous inner one. The black dashed lines indicate reference circles. The yellow pentagram denotes the location of the Sun.}
\label{fig:twisted_pattern}
\end{figure*}

We adopt the radial density profile to be a fixed triple power-law model, with the parameters $r_{b,1}$, $r_{b,2}$, $\alpha_{1}$, $\alpha_{2}$, $\alpha_{3}$ fixed to best-fit values from Section~\ref{sec:triple power-law model results}, and then introduce the dependence of the shape and orientation parameters on radius (see Equation~\ref{equ:variable_model}).

\begin{figure}
\begin{center}
\includegraphics[width=0.45\textwidth]{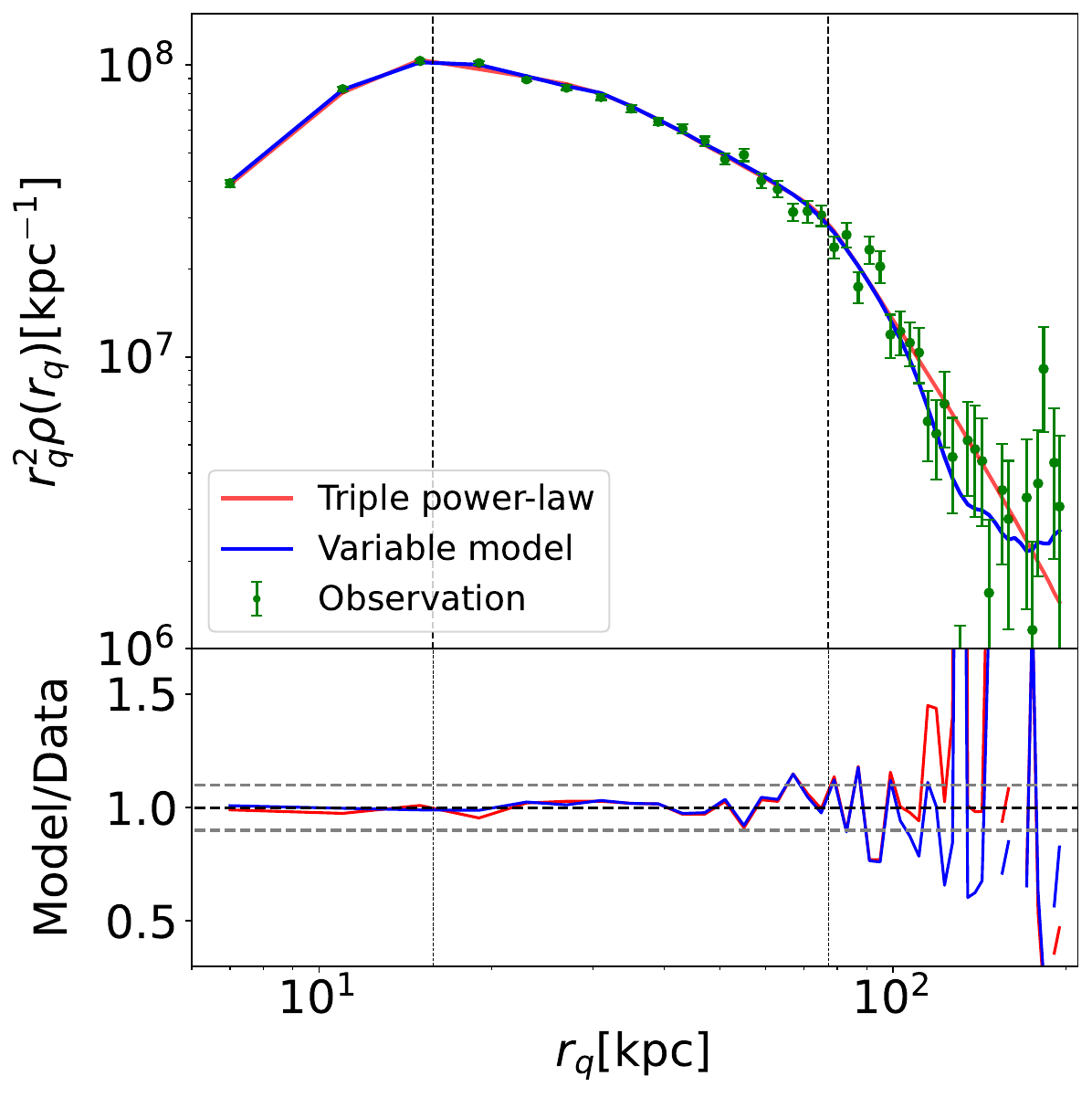}%
\end{center}
\caption{Similar to Figure~\ref{fig:1rb_2rb_compare}, but the blue solid curve shows the variable model, by allowing the orientation angles and axis ratios to vary with radius. The red solid curve and green dots with errorbars are the same as those in Figure~\ref{fig:1rb_2rb_compare}, which are the best-fit triple power-law model profile and the real data. The model predictions have been convolved with the angular and radial selection functions (see Section~\ref{sec:selection function}), and have been renormalized to give the same total number of stars as real data.}
\label{fig:2rb_variable_compare}
\end{figure}

\begin{figure*}
\begin{center}
\includegraphics[width=0.7\textwidth]{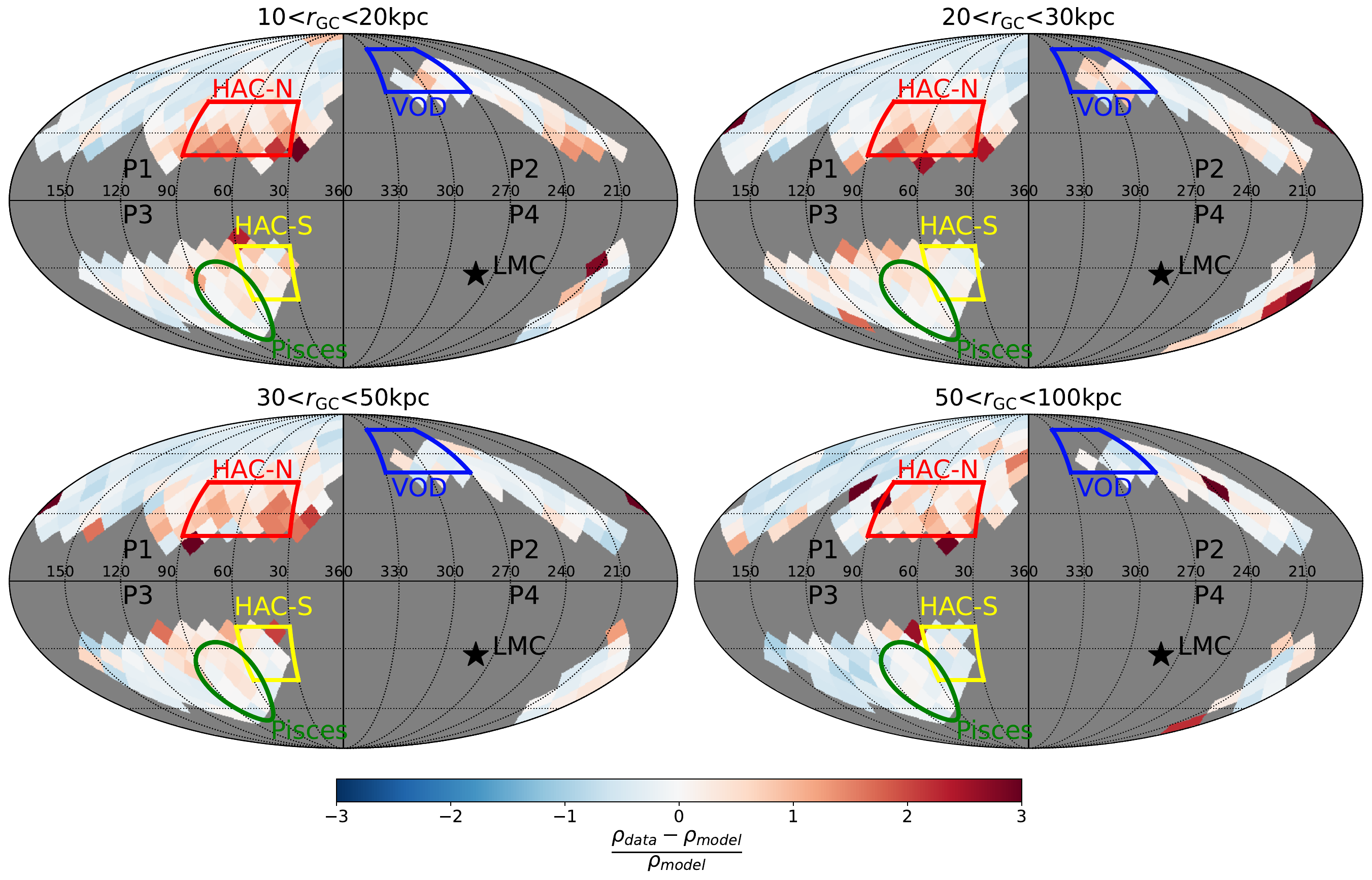}%
\end{center}
\caption{Angular residuals between data and the best-fit variable model in Mollweide projection in Galactic coordinates. Here, selection effects have been corrected. Four panels show the residuals in four different galactocentric distance bins of $10~\mathrm{kpc}<r_\mathrm{GC}<20~\mathrm{kpc}$, $20~\mathrm{kpc}<r_\mathrm{GC}<30~\mathrm{kpc}$, $30~\mathrm{kpc}<r_\mathrm{GC}<50~\mathrm{kpc}$ and $50~\mathrm{kpc}<r_\mathrm{GC}$ (see the text on top). The black pentagram in each panel marks the current location of LMC. The HAC-S (yellow), HAC-N (red), Pisces (green), and VOD (blue) regions are enclosed by the corresponding colored solid lines. The definition of these regions can be found in Table~\ref{table:regions}.}
\label{fig:angular_residual}
\end{figure*}

\begin{figure*}
\begin{center}
\includegraphics[width=0.7\textwidth]{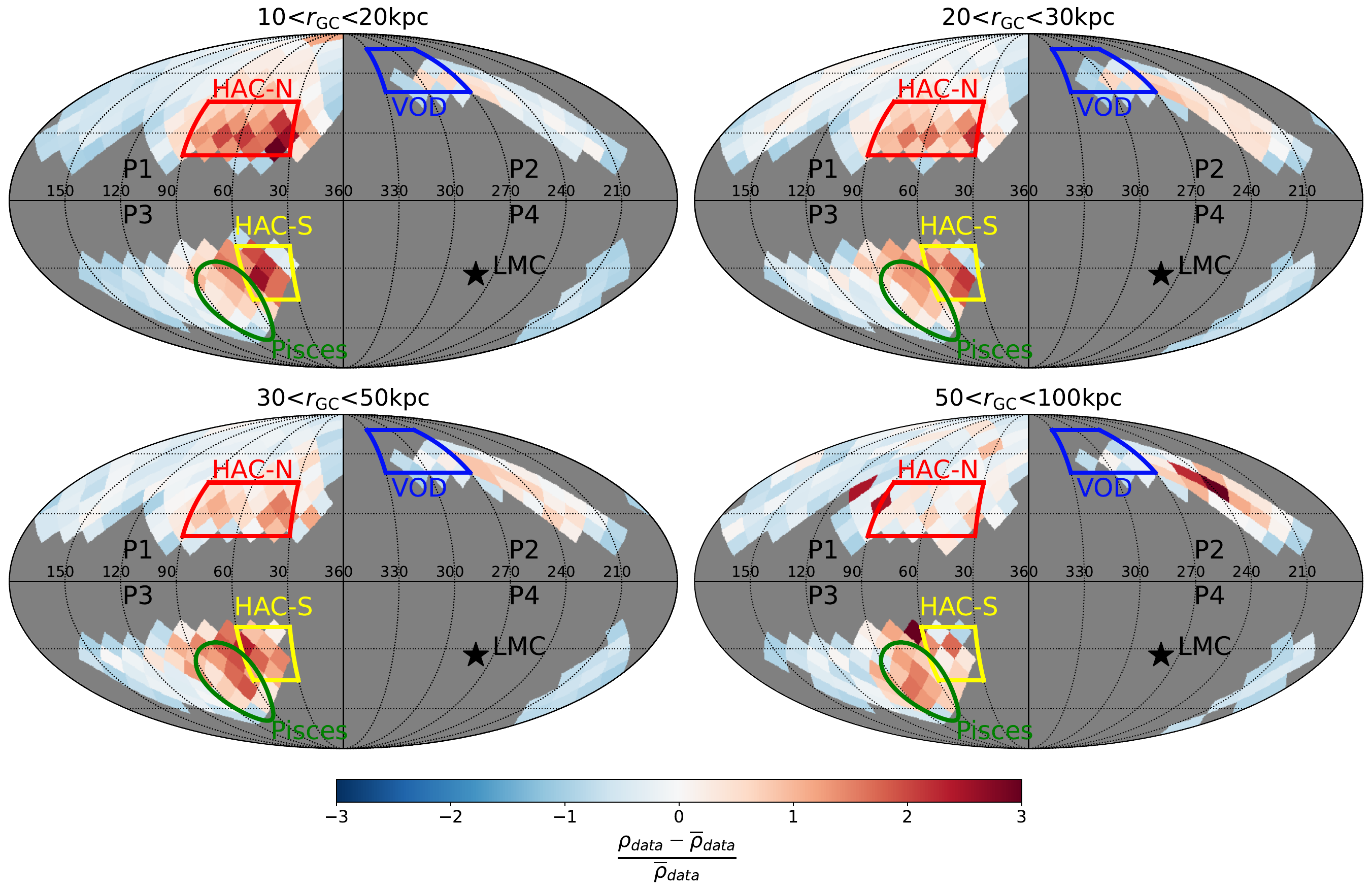}%
\end{center}
\caption{Density contrast of  MW stellar halo in Mollweide projection in Galactic coordinates. Here, selection effects have been corrected. The figure is made in four different bins of galactocentric distances $10~\mathrm{kpc}<r_\mathrm{GC}<20~\mathrm{kpc}$, $20~\mathrm{kpc}<r_\mathrm{GC}<30~\mathrm{kpc}$, $30~\mathrm{kpc}<r_\mathrm{GC}<50~\mathrm{kpc}$ and $50~\mathrm{kpc}<r_\mathrm{GC}<100~\mathrm{kpc}$ (see the text on top of each panel). The density contrast is defined as the difference between observed density at different sky pointings and the mean observed density at a given $r_\mathrm{GC}$ range, further divided by the mean density. The black pentagram in each panel marks the current location of LMC. The HAC-S (yellow), HAC-N (red), Pisces (green), and VOD (blue) regions are marked by the corresponding colored solid lines. The definition of these regions can be found in Table~\ref{table:regions}.}
\label{fig:obs_angular_residual}
\end{figure*}

Figure~\ref{fig:mcmc_variable} shows the posterior contours for the variable model. Different orders of $k_{\phi i}$/$k_{\theta i}$ are highly correlated with each other, whereas the other model parameters are more independently constrained. In the top right corner of Figure~\ref{fig:mcmc_variable}, we present two panels which show the dependence of rotation angles (top panel) and flattenings (bottom panel) on $r_\mathrm{GC}$. The thick lines are the prediction from the best-fit model, and the shaded regions present the 1-$\sigma$ uncertainty of the variable model.

We find that both the pitch and yaw angles, $\theta$ and $\phi$, decrease with increasing $r_\mathrm{GC}$. The yaw angle is more aligned with the line connecting the Sun and Galactic center within 20~kpc, which increases to larger absolute values at larger distances. Interestingly, there is a clear trend that the inner stellar halo within $r_\mathrm{GC}\lesssim30$~kpc is more aligned with the disk, with the absolute value of the pitch angle, $\theta$, smaller than 45~$\degree$. However, the absolute value of $\theta$ becomes greater than 45~$\degree$ beyond 30~kpc, indicating that the major axis of the outer stellar halo becomes more perpendicular to the MW disc, i.e., the orientation of the stellar halo flips with distance. Here, a negative pitch angle means that the major axis of the stellar halo is pointing to the South Galactic Pole. The pitch angle even reaches ~$-$80$\degree$ when $r_\mathrm{GC}$ is around 60~$\mathrm{kpc}$.

Moreover, the flattening parameter, $q$, of the minor axis is first increasing with the radius, reaching a maximum at $r_\mathrm{GC}\sim45$~kpc and then decreasing again. The flattening of the intermediate axis, $p$, keeps decreasing. In fact, the stellar halo is more prolate at $r_\mathrm{GC}>30$~kpc, where $p$ and $q$ are closer to each other, while both are smaller than 1. At a smaller $r_\mathrm{GC}$, $p$ is closer to 1, whereas $q$ is smaller by $\sim$0.15, indicating the more nearby inner stellar halo is more oblate. 

Figure~\ref{fig:twisted_pattern} more intuitively shows how the best-fit variable ellipsoidal model changes with radius in galactocentric Cartesian coordinates. The red solid lines are isodensity contours. The black dashed lines indicate reference circles.
The left and middle panels show the density distributions in the $x$-$z$ ($y=0$) and $y$-$z$ ($x=0$) slices that are perpendicular to the disk. The stellar halo exhibits a pronounced twist: the isodensity contours systematically rotate towards perpendicular to the disk plane, and the flattening changes with increasing galactocentric distance. The right panel shows the density distribution in the disk plane ($z=0$). We also see a trend that on this plane, the stellar halo is closer to circles at smaller radii, which becomes more elliptical with the increase in radius.

Combining the trends in orientation angles and the two axis ratios with $r_\mathrm{GC}$, it is straightforward to see that the inner stellar halo within 30~kpc is oblate and the major axis is more aligned with the disk (with the minor axis perpendicular to the disk). On the other hand, the more distant outer stellar halo becomes more prolate and perpendicular to the disk, with the major axis more aligned with the Vast Polar Structure of MW satellites, GCs, and stellar streams \citep[e.g.][]{2012MNRAS.423.1109P,2014ApJ...790...74P}. However, the trends of flattening parameter and orientation angles at a larger distance ($r_\mathrm{GC}>\sim50-60$~kpc) are still very uncertain due to quite large model uncertainties. 

Similar to Figure~\ref{fig:1rb_2rb_compare}, Figure~\ref{fig:2rb_variable_compare} presents the comparison between the observation, the best-fit triple power-law model, and the variable model profile, as a function of the flattened radius, $r_q$. The two vertical black dashed lines present two break radii from the best-fit triple power-law model in the previous subsection. Both the triple power-law model and variable model fit the 1-dimensional density profile well at $r_{b,1}<r_q<r_{b,2}$, while the variable model fits the observation better within the first break radius. It is hard to distinguish which model fits the observation better at $r_q>r_{b,2}$ due to the much smaller number of stars there and the large error bars. Nevertheless, the difference of the AIC values between the variable model and the triple power-law model is about $-$136, which means that the data prefers the variable model. We have further checked the angular residuals of the two models and found that the variable model has smaller residuals.

Figure~\ref{fig:angular_residual} shows the angular residuals of the densities predicted by the variable model in Mollweide projection in Galactic coordinates and in four different ranges of $r_\mathrm{GC}$ (see the text on top of each panel). The density at a given pixel is the ratio between the number of stars at this pixel and the corresponding volume.
The angular residual is further defined by the difference between observed density and predicted density at different $l$ and $b$, divided by the prediction at a given distance range. Here, the LMC is marked by a black solid star.
However, we find that, even with the most flexible variable model, we still can not fully capture all angular substructures of the real stellar halo. 

There are some overdense regions in Figure~\ref{fig:angular_residual}. For example, we find that the region enclosed by the red solid line presents some over-density in all four panels at different distances. We also observe a mild over-density of the areas enclosed by the green, blue, and yellow solid lines, although it is not present in every panel. The enclosed regions by the red, yellow, and blue solid lines are associated with three well-known overdense regions, HAC-N, HAC-S and VOD \citep{2001ApJ...554L..33V,2002ApJ...569..245N,2007ApJ...657L..89B,2008ApJ...673..864J,2009MNRAS.398.1757W,2010ApJ...708..717S,2014MNRAS.440..161S}, which are chemodynamically consistent with GSE stars \citep{2019MNRAS.482..921S,2021ApJ...923...92N,2022ApJ...936L...2P}. Besides, the region enclosed by the green line contains Pisces overdensity \citep{2019MNRAS.488L..47B,2021Natur.592..534C}, which could be associated with the transient density wake of LMC. The definition of these solid lines enclosed regions can be found in Table~\ref{table:regions}. We also see some slightly underdense regions in each panel. These slightly underdense regions might be caused by the existence of the overdense regions we talked about before, and thus the model tends to be shifted to larger values in order to better match the few overdense regions.
We further explore these substructures in Section~\ref{sec:Anisotropic stellar halo density distribution and the LMC wake}.

\begin{figure*}
\begin{center}
\includegraphics[width=1\textwidth]{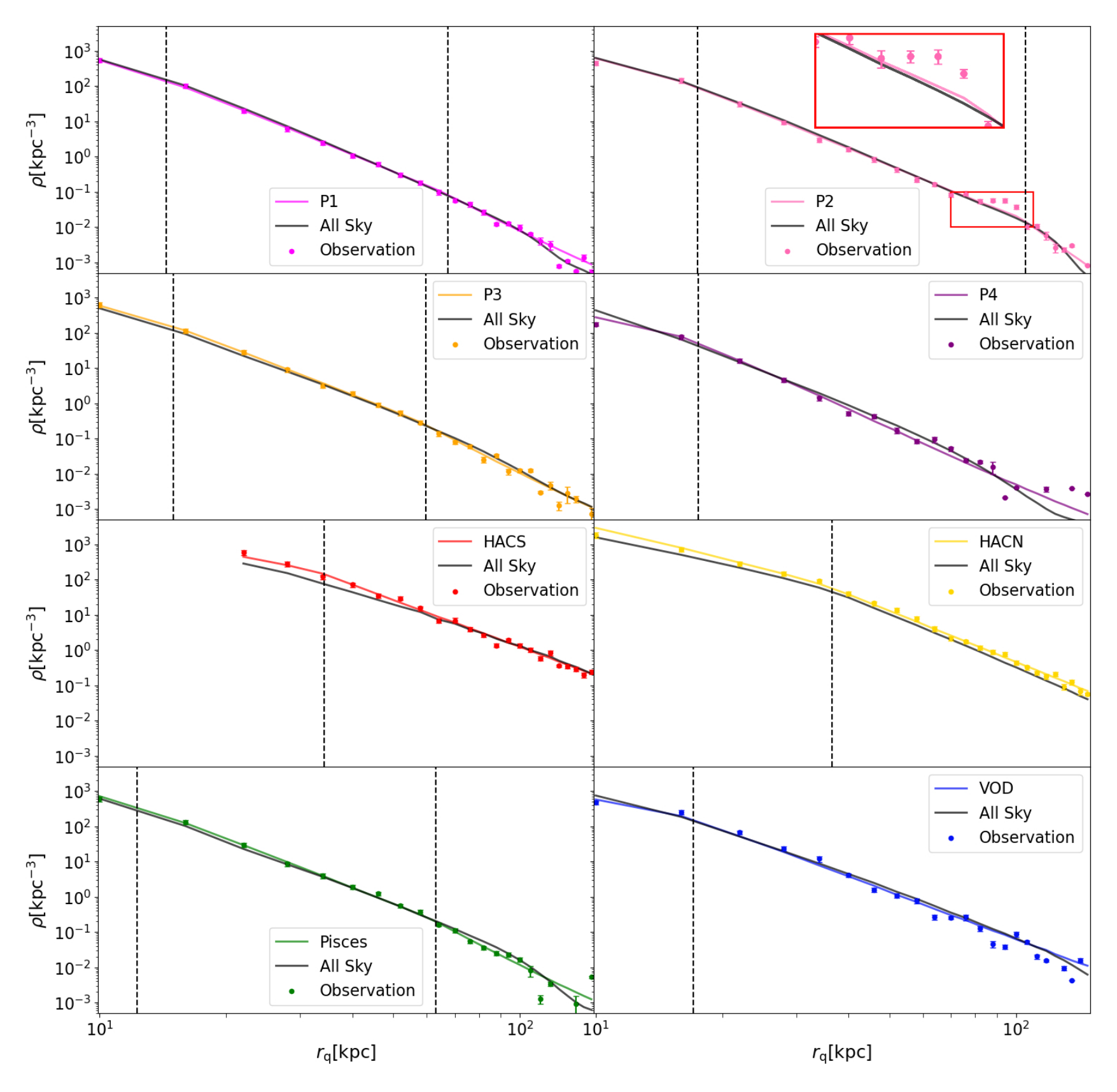}%
\end{center}
\caption{In each panel, we compare the predicted density profiles of the variable model from Section~\ref{sec:variable results} in the corresponding sky position (black curve) with the result of the triple power-law model (different color curves, see the legend), which directly fits this region. The dashed lines stand for the break radii of the density profile in this region.
Note that although the black curves are based on the global fitting with the full data, they differ across different panels, as each panel corresponds to a particular pointing. Selection effects have been corrected in this Figure. All colored model density profiles have been renormalized to give the same number of stars as real data at the given sky position after correcting selection effect. The black curve is renormalized to the whole sample.}
\label{fig:radial_profile_different_position}
\end{figure*}

\subsection{Spatial Anisotropies of Stellar Halo Density Distribution and the LMC Density Wake}
\label{sec:Anisotropic stellar halo density distribution and the LMC wake}

In Sections~\ref{sec:triple power-law model results} and \ref{sec:variable results}, we fit the stellar halo with a continuous and smooth ellipsoid model. However, our MW stellar halo is full of substructures, and even the variable model cannot fully capture these density fluctuations. Thus, we show the observed density contrast in Figure~\ref{fig:obs_angular_residual} in the Mollweide projection in Galactic coordinates. The density contrast is defined by the difference between the observed density at given ($l$,$b$) and the mean observed density in the ellipsoidal shell at the given distance, and is divided by the latter. To get rid of any artificial over- or underdense features, we have corrected the incompleteness with angular and radial selection functions (Equation~\ref{equ:selection function1} and Equation~\ref{equ:selection function2}). There are a few overdense regions, marked by colored solid lines, which are roughly revealed in Figure~\ref{fig:angular_residual} above. In addition to these overdense regions, there is an overdense region in the right top part of the sky ($210\degree<l<330\degree$ and $b>0\degree$), with the over-density becoming more prominent with the increase in distances.

For these overdense regions, the stellar halo may have different break radii and power-law slopes. We then divide the footprint into a few pieces and model the shape, orientation, and radial density profile separately. We explore the anisotropy in eight regions defined in Table~\ref{table:regions} with the triple power-law model. 
Here, HAC-N\footnote{Note that the area of HAC-N defined in this work is twice as large as previous studies \citep{joao}. We have further checked that the area ($30\degree<l<60\degree, -45\degree<b<-20\degree$) and HAC-N defined in this work share similar radial density profile, shape, and orientation parameters by forward modelling. Hence, we believe that a different definition of HAC-N would not affect the results and keep the definition of HAC-N in the rest of this work.}, HAC-S, and VOD are spatial overdensities observed in the sky and chemodynamically associated with the GSE debris \citep{2019MNRAS.482..921S,2022ApJ...936L...2P}. Pisces overdensity is correlated with LMC\footnote{\cite{2019ApJ...884...51G} presented high-resolution N-body simulations of the MW-LMC interaction. They identified two main density wakes, the transient wake and the collective wake in the MW stellar halo induced by the LMC. The transient wake is a trailing over-density following the orbit of the LMC, while the collective wake is a broader, persistent over-density in the northern halo caused by resonant global response.} \citep{2019ApJ...884...51G,2021Natur.592..534C,70kpc_halo}. 
P1, P2, P3, and P4 are the first, second, third, and fourth quadrants in the galactic projection. 
P2 is an overdense region, while P4 is an underdense region. These are likely associated with the global collective response of the stellar halo to LMC infall, and we will discuss them later in this section.

Figure~\ref{fig:radial_profile_different_position} presents the comparison between the global best-fit variable model obtained in Section~\ref{sec:variable results} and the corresponding triple power-law models (different color curves in each panel) that directly fit the measured density profile in each region (see the legend).
The global variable model profiles differ in different panels because the variable model is dependent on angular position. 
Here, we have corrected the selection effect. In all panels of Figure~\ref{fig:radial_profile_different_position}, the triple power-law model at a given sky position fits the observed density profile well. We also find that the variable model does not present a very large difference compared with the best-fit triple power-law model in each region, which indicates that the variable model is flexible in terms of capturing the main density variations over the sky. 
However, in the HAC-N region, observation points are slightly higher than the global black curve at all distances, while in the HAC-S and Pisces regions, the observation points are slightly higher than the black curve at smaller distances. And in the P4 region, the variable model slightly overestimates the density profile. These are all consistent with Figure~\ref{fig:angular_residual}.

Table~\ref{table:regions} provides the best-fit parameters in each panel of Figure~\ref{fig:radial_profile_different_position}. In particular, we find that for the best-fit triple power-law model of HAC-S, HAC-N, VOD, and P4 regions, the difference between the best-fit second power-law slope $\alpha_2$ and the third power-law slope $\alpha_3$ is negligible when compared with the uncertainties of $\alpha_2$ and $\alpha_3$. Hence, the triple power-law model degrades to a double power-law model, and thus we omit the second break radius $r_{b,2}$ and $\alpha_3$ in these regions. The break radius of VOD is about 15~kpc, which is consistent with the first break radius we find in Section~\ref{sec:triple power-law model results}. 
The break radii of both HAC-N and HAC-S are around 30~kpc, which is consistent with the second break radius reported in \citep{Han_stellar_halo_density_profile} for GSE. This may indicate that HAC-N and HAC-S are highly correlated with GSE stars \citep{2019MNRAS.482..921S,2022ApJ...936L...2P,2024MNRAS.532.2584Y}. Besides, we find that the ellipsoids of HAC-N and HAC-S are highly prolate and present very large pitch angles. We report a pitch angle of 100$\degree$ and 53$\degree$ for HAC-S and HAC-N, respectively. However, given the small area of these overdense regions, the best-constrained axis ratios, pitch, and yaw angles could be subject to large uncertainties due to extrapolations, though the associated statistical errors are small, and thus we do not over-interpret these shape and position parameters. Moreover, we have tried directly fitting spherical models. The best-fit density slope parameters and break radii remain consistent with those of the ellipsoidal models adopted here.

\begin{table*}[ht]
\centering
\caption{Best-fit parameters of triple power-law model in different regions.}
\renewcommand{\arraystretch}{1.8}
\scriptsize      
\setlength{\tabcolsep}{3.5pt}  % 调紧列间距
\begin{tabular}{@{}lcccccccccc@{}}
\toprule
Name & $(l, b)$ & $r_{b,1}$ [kpc] & $r_{b,2}$ [kpc] & $\alpha_{1}$ & $\alpha_{2}$ & $\alpha_{3}$ & $p$ & $q$ & $\theta$ [$^\circ$] & $\phi$ [$^\circ$] \\
\midrule
Halo & $0^\circ<l<360^\circ, -90^\circ<b<90^\circ$ & $16.0^{+0.25}_{-0.20}$ & $76.3^{+2.37}_{-1.94}$ & $1.50^{+0.07}_{-0.07}$ & $3.45^{+0.02}_{-0.02}$ & $5.20^{+0.14}_{-0.14}$ & $0.85^{+0.01}_{-0.01}$ & $0.74^{+0.01}_{-0.01}$ & $-43.8^{+0.7}_{-0.7}$ & $-26.9^{+1.0}_{-1.0}$ \\
P1 & $0^\circ<l<180^\circ, 0^\circ<b<90^\circ$ & $14.4^{+0.2}_{-0.2}$ & $69.3^{+3.3}_{-3.5}$ & $1.28^{+0.07}_{-0.08}$ & $3.48^{+0.02}_{-0.02}$ & $4.84^{+0.17}_{-0.16}$ & $0.93^{+0.01}_{-0.01}$ & $0.71^{+0.01}_{-0.01}$ & $-38.6^{+0.8}_{-0.8}$ & $-0.4^{+0.3}_{-0.6}$ \\
P2 & $180^\circ<l<360^\circ, 0^\circ<b<90^\circ$ & $17.7^{+0.7}_{-1.0}$ & $103.1^{+2.9}_{-2.2}$ & $1.72^{+0.17}_{-0.33}$ & $3.56^{+0.03}_{-0.02}$ & $7.47^{+0.47}_{-0.15}$ & $0.71^{+0.08}_{-0.07}$ & $0.66^{+0.03}_{-0.03}$ & $-52.9^{+1.0}_{-1.0}$ & $-14.2^{+8.0}_{-8.0}$ \\
P3 & $0^\circ<l<180^\circ, -90^\circ<b<0^\circ$ & $14.9^{+0.5}_{-0.5}$ & $60.0^{+3.0}_{-0.9}$ & $1.34^{+0.15}_{-0.15}$ & $3.25^{+0.03}_{-0.03}$ & $4.89^{+0.15}_{-0.15}$ & $0.66^{+0.01}_{-0.01}$ & $0.64^{+0.01}_{-0.01}$ & $-97.0^{+0.7}_{-0.7}$ & $-30.3^{+1.6}_{-1.6}$ \\
P4 & $180^\circ<l<360^\circ, -90^\circ<b<0^\circ$ & $17.9^{+1.5}_{-10.6}$ & & $0.91^{+1.57}_{-0.39}$ & $3.89^{+0.16}_{-1.51}$ & & $0.77^{+0.04}_{-0.08}$ & $0.73^{+0.06}_{-0.03}$ & $-5.8^{+3.4}_{-4.5}$ & $-1.6^{+1.3}_{-2.6}$ \\
Pisces & $(l-74^\circ)^2+(b+47^\circ)^2<400$ & $9.5^{+5.1}_{-4.5}$ & $60.1^{+5.3}_{-10.1}$ & $1.95^{+1.10}_{-0.99}$ & $3.22^{+0.06}_{-0.04}$ & $4.99^{+0.25}_{-0.18}$ & $0.74^{+0.04}_{-0.03}$ & $0.72^{+0.06}_{-0.03}$ & $-76.1^{+3.8}_{-3.1}$ & $-21.6^{+10.0}_{-63.6}$ \\
HAC-S & $30^\circ<l<60^\circ, -45^\circ<b<-20^\circ$ & $34.3^{+5.8}_{-8.1}$ & & $0.52^{+0.31}_{-0.48}$ & $2.63^{+0.05}_{-0.03}$ & & $0.21^{+0.06}_{-0.02}$ & $0.17^{+0.05}_{-0.03}$ & $-100.1^{+1.4}_{-1.2}$ & $-0.3^{+0.3}_{-0.44}$ \\
HAC-N & $30^\circ<l<90^\circ, 20^\circ<b<45^\circ$ & $31.8^{+9.5}_{-6.4}$ & & $1.71^{+0.09}_{-0.15}$ & $2.98^{+0.06}_{-0.06}$ & & $0.32^{+0.09}_{-0.06}$ & $0.31^{+0.09}_{-0.06}$ & $-52.8^{+2.8}_{-2.7}$ & $-0.9^{+0.6}_{-1.1}$ \\
VOD & $270^\circ<l<330^\circ, 50^\circ<b<75^\circ$ & $15.8^{+8.6}_{-2.6}$ & & $1.42^{+0.98}_{-1.06}$ & $3.15^{+0.27}_{-0.22}$ & & $0.69^{+0.18}_{-0.21}$ & $0.59^{+0.10}_{-0.09}$ & $-55.1^{+13.8}_{-9.9}$ & $-10.1^{+8.2}_{-20.5}$ \\
\bottomrule
\end{tabular}
\label{table:regions}
\end{table*}

For P1 and P3 regions, which contain HAC-N and HAC-S, respectively, the best-fit radial density profile is a triple power-law model. P3 is nearly prolate, while P1 is more likely to be oblate. P1 and P3 share similar break radii, with $r_{b,1}$ around 14~kpc and $r_{b,2}$ around 60 to 70~kpc, which is consistent with Section~\ref{sec:triple power-law model results}.  We also see a second break radius around 60~kpc in the Pisces region, which is consistent with \cite{70kpc_halo}, who measured the radial density of the Pisces region based on MSTO stars and found a break radius at $\sim$70 kpc. P3 shares a similar second break radius with Pisces, which is likely because the P3 region is dominated by Pisces at larger distances. N-body models \citep[e.g.][]{2019ApJ...884...51G,2021Natur.592..534C,2024MNRAS.534.2694S,2025arXiv250703663S,2025ApJ...988..156C} predict that the first infall of the LMC produces an over-density at $\sim$60 to 70~kpc in the Pisces region, and thus the second break radius we found in the Pisces might be related to the LMC transient wake.

For P2 and P4 regions, the best-fit density profiles deviate from those in P1 and P3 regions. We find that the second break radius of P2 is approximately 100~kpc. Moreover, we see a clear over-density of the observational dots, peaked at 90~kpc, as marked by the red triangular box with the zoom-in figure shown in the bottom left of the P2 panel. The obvious over-density and a break radius at 90-100~kpc may be new evidences of the LMC collective wake. Moreover, the P2 region in the northern sky is denser than the underdense P3 region by a factor of 4.5 between 50 and 100~kpc. This is also related to the global collective response to the LMC infall, which results in over and under-density in the northern and southern sky as predicted by numerical simulations \citep[e.g.][]{2019ApJ...884...51G,2021Natur.592..534C,2024MNRAS.534.2694S,2025arXiv250703663S}. With the large sample of DESI K giant stars extending to large distances, we have clearly detected this collective density wake, consistent with theoretical predictions.

In the end, we emphasize that though numerical simulations have predicted north-south asymmetry that also exists in P1 and P3 regions, i.e., P1 is predicted to be an overdense region while P3 is an underdense region, we fail to see this in the data. If excluding the HAC-N, HAC-S, and Pisces overdense regions, the other parts of P1 and P3 show very similar densities. The inconsistency between simulation predictions and real observations is at least partially related to how star particles are incorporated in N-body simulations of the MW-LMC encounter. Almost all such simulations that predicted the density wakes in the past do not contain star particles. Instead, star particles are often created by tagging dark matter particles following, for example, the method of \cite{2013MNRAS.435..901L}, assuming star and dark matter particles are in equilibrium. In particular, dark matter particles are tagged as stars based on a weight calculated from the distribution function obtained through Eddington inversion and under spherical assumptions. These assumptions may not hold solid for the real MW stellar and dark halos, and thus may be responsible for part of the discrepancies. In fact, simulation studies do reveal the spatial and kinematic biases of the stellar halo with respect to the dark halo \citep[e.g.][]{2024ApJ...976..187H}.

LMC also has a significant effect on changing the velocity pattern of the MW outer stellar halo. The velocity imprints of LMC on the MW stellar halo have been studied using BHBs from DESI \citep{2025MNRAS.542..560B}. Since the main science of this paper is to look at the density distribution of the stellar halo, we leave more detailed studies on the velocity part in a follow-up study (Li et al., in prep.).

\subsection{Model Dependence on Metallicity}
\label{sec:Model dependence on Metallicity}
In this section, we explore the model dependence on metallicity, [Fe/H]. We divide the K giant sample into seven subsamples by [Fe/H]: $-1.1<\mathrm{[Fe/H]}<-0.85$, $-1.2<\mathrm{[Fe/H]}<-0.9$, $-1.3<\mathrm{[Fe/H]}<-1.0$, $-1.4<\mathrm{[Fe/H]}<-1.1$, $-1.6<\mathrm{[Fe/H]}<-1.2$, $-1.9<\mathrm{[Fe/H]}<-1.3$, $-5<\mathrm{[Fe/H]}<-1.4$. Each subsample contains at least 8,000 stars and can overlap with others in metallicity to make sure that each subsample contains enough stars. We adopt the triple power-law model and then run MCMC and get the best-fit parameters of each subsample.

Figure~\ref{fig:parameters_vs_feh} shows the dependence of best-fit model parameters on [Fe/H]. The points are the best-fit parameters as a function of the median [Fe/H] in each subsample. The black horizontal solid line shows the best-fit parameters in Figure~\ref{fig:mcmc_2rb}, i.e., for the full sample, and the gray shaded region presents the uncertainties. In this Figure, we can see some interesting trends in different panels:

\begin{figure}
\begin{center}
\includegraphics[width=0.4\textwidth]{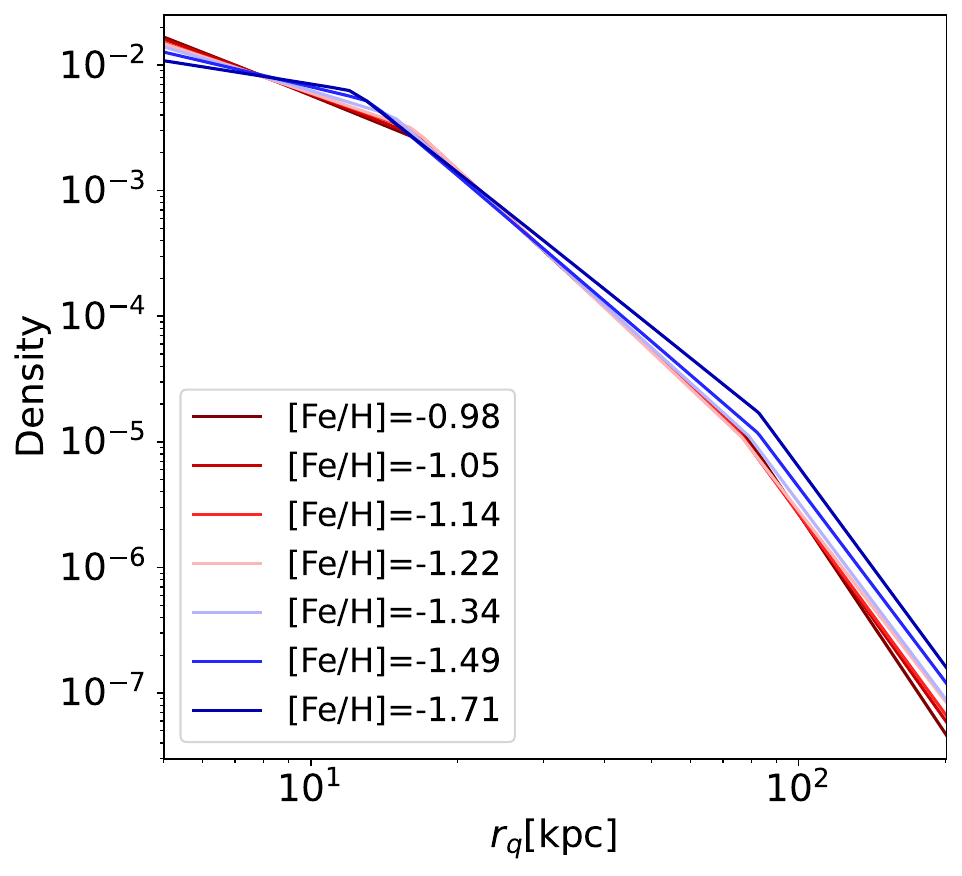}%
\end{center}
\caption{The best-fit selection effect free radial density profiles for subsamples with different metallicity (see the legend). All profiles have been renormalized to unity.}
\label{fig:feh_vs_profile}
\end{figure}

We find that the first break radius $r_{b,1}$ generally decreases with the decrease in [Fe/H], while the second break radius $r_{b,2}$ shows a more complex trend, increasing only for the two most metal-poor subsamples. Additionally, the power-law index $\alpha_{1}$ is decreasing with the decrease in [Fe/H], while $\alpha_2$ and $\alpha_3$ exhibit non-monotonic behaviors with metallicity, although there is a general trend of decreasing. These trends suggest that the density profile of the stellar halo becomes shallower for more metal-poor populations, but with notable exceptions and variations among the sub-samples.

We also see some features in the orientation and shape parameters. The two flattening parameters $p$ and $q$ are slightly increasing for more metal-poor populations. The rotation angles $\theta$ and $\phi$ also become more positive for the more metal-poor population. However, the fractional changes in $p$, $q$, and $\theta$ are only a few percent. $\phi$ almost does not change at $\mathrm{[Fe/H]}>-1.4\,\mathrm{dex}$, but shows some more prominent change for the last two metal-poor bins. 

Figure~\ref{fig:feh_vs_profile} shows a direct comparison among the best-fit model density profiles for subsamples with different metallicities. The model density profiles have been renormalized to unity.
% It more clearly demonstrates the difference among halo stars. 
It is clear that with the decrease in [Fe/H], the distribution of halo stars becomes more extended. 
Since the break radius is likely correlated with an accretion event \citep[e.g.][]{2009MNRAS.398.1757W,2013ApJ...763..113D,Hernitschek_2018,2021ApJ...923...92N,Han_stellar_halo_density_profile}, the dependence of $r_{b,1}$/$r_{b,2}$ may reflect a time-dependent stripping process of a dwarf galaxy to form the outer and inner stellar halos at different times. 
In general, it is expected that the more metal-poor outskirts of a dwarf galaxy are stripped earlier, at larger distances from the center of the host, in contrast to the more metal-rich inner regions, which are later disrupted/stripped at closer distances \citep{2020A&A...642L..18K,2022ApJ...937...12A}.
% Usually, the more metal-poor outskirts of a dwarf galaxy are stripped earlier, when the dwarf galaxy is at larger distances from the host center. The more metal-rich inner region of a dwarf galaxy is likely stripped later when the dwarf galaxy is closer to the host center. 
Thus, the more metal-poor stellar halo tends to be more extended.

Based on Figures~\ref{fig:parameters_vs_feh} and \ref{fig:feh_vs_profile}, we conclude that the radial density profile has a clear dependence on metallicity, showing that the metal-poor stellar halo is more extended, whereas the dependence of shape and orientation parameters on metallicity also exists, but is weaker.

\section{Discussions}
\label{sec:discussions}

\subsection{Physical Interpretations on the Twisted Inner and Outer Stellar Halos}

Our results reveal that the inner ($\lesssim30$~kpc) and outer stellar halos of our MW have different shapes and orientations. The inner stellar halo is oblate and more aligned with the disk, whereas the outer stellar halo is more prolate and perpendicular to the disk. Our finding here is likely related to the Vast Polar Structure (VPOS) of our MW \citep{2012MNRAS.423.1109P,2014ApJ...790...74P}, that about 11 classical MW satellites are moving on orbits perpendicular to the MW disk. 
Nowadays, a growing number of studies argue that groups of satellites moving coherently do exist in $\Lambda$CDM simulations, which may form from the same group of galaxies \citep[e.g.][]{2015MNRAS.452.3838C,2015MNRAS.449.2576C,2018MNRAS.476.1796S,2019MNRAS.488.1166S,2023NatAs...7..481S}.

Nevertheless, the VPOS indeed shows a group of satellites having high inclination orbits perpendicular to the disk. Now our work here also shows that the outer stellar halo is more prolate and perpendicular to the disk. Hence the orientation of MW satellites and the outer stellar halo is more consistent, and the MW disk plus inner halo seems to be misaligned with the matter distribution in the outskirts. 
Thus, the key is to explain the misalignment between the inner and outer matter distribution of our MW. 

First of all, in numerical simulations, the existence of baryons is predicted to affect the angular momentum of the inner dark matter halo \citep{2021ApJ...913...36E,2022MNRAS.515.2681C}. \cite{2010MNRAS.404.1137B} pointed out that the formation of the galaxy spins up the dark matter within 0.1 times the virial radius, such that the specific halo angular momentum increases by $\sim$50\% in the median. This results in a better alignment between the angular momenta of the central galaxy and the inner host dark halo. Though we measure the shape and orientation of the MW stellar halo, we may expect to see similar trends between the stellar and dark halos.

Another possible scenario could be related to the change or flip in the MW disk and inner halo angular momentum. It was shown by \cite{2017MNRAS.472.3722G} through cosmological hydrodynamical simulation that after the infall of a satellite galaxy, the angular momentum of the satellite and the host galaxy disk would align. This is not only because the orbital plane of the infalling satellite is affected by the disk potential, but also because the disk is responding to the infalling massive satellite. Moreover, studies based on numerical simulations have reported that the spin vector of the inner halo experiences much more frequent flips than the halo as a whole \citep{2012MNRAS.420.3324B}, and a significant fraction of halos having had a large spin flip are associated with minor mergers \citep{2016MNRAS.461.1338B}. If the above scenario is also true for our MW, it indicates that our MW disk plane and the inner stellar halo may have flipped by $\sim$90~degrees, due to the infall of a massive satellite. Note that the more frequent flips in the central galaxy and inner halo are likely due to their shorter dynamical time scales than the outer halos.

Observational constraints on the MW dark halo shape often rely on stellar streams or halo stars spanning wide ranges of sky footprint. Through the modeling of stellar streams \citep{2025ApJ...985L..22N,2025MNRAS.539.2718P}, \cite{2023MNRAS.524.2124P} also identified slightly prolate MW dark halos, with their major axis more perpendicular to the disk.
Very recently, starting from the assumption that given a correct potential model, the distribution function of MW halo stars does not evolve with time \citep[e.g.][]{2016MNRAS.456.1017H,2016MNRAS.456.1003H,2025MNRAS.538.1442L}, and with the model extended to non-spherical case \citep{2025A&A...703A..43Z}, \cite{2025arXiv251008684Z} reported that the MW dark halo within 50~kpc is nearly oblate, with the long-intermediate axis plane of the dark-matter halo perpendicular to the Galactic disk beyond 20~kpc. Similar conclusions are reported of simulated MW-like systems \citep[e.g.][]{2021MNRAS.504.6033S}. All these studies are showing a broadly consistent picture of our measured MW inner and outer stellar halo orientations. 

In the end, we would like to emphasize that there is a large scatter in the alignment angles between the spins or major axes of galaxies and host dark haloes, with a median mis-alignment angle of about 30 degrees as reported in both simulations \citep{2010MNRAS.404.1137B} and real observations by calculating the galaxy shape-shape correlations and shape-shear correlations \citep[e.g.][]{Okumura2009,Okumura2009b,2023ApJ...954....2X,2023ApJ...957...45X}. Moreover, the correlation strength is much weaker in disk galaxies than in elliptical galaxies \citep[e.g.][]{2015SSRv..193..139K,2023ApJ...954....2X}. This is likely because galaxy disks are mainly formed from the accreted gas during a certain period of time, whereas the spin of the outer halo is formed by later accreted material, which does not necessarily have the same spin as early accreted gas.

\subsection{Comparison with Other Studies}
Finally, we present the comparison of the radial density profile between this work and the literature, and give a brief discussion on the tension in radial density profiles.

\begin{figure}
\begin{center}
\includegraphics[width=0.49\textwidth]{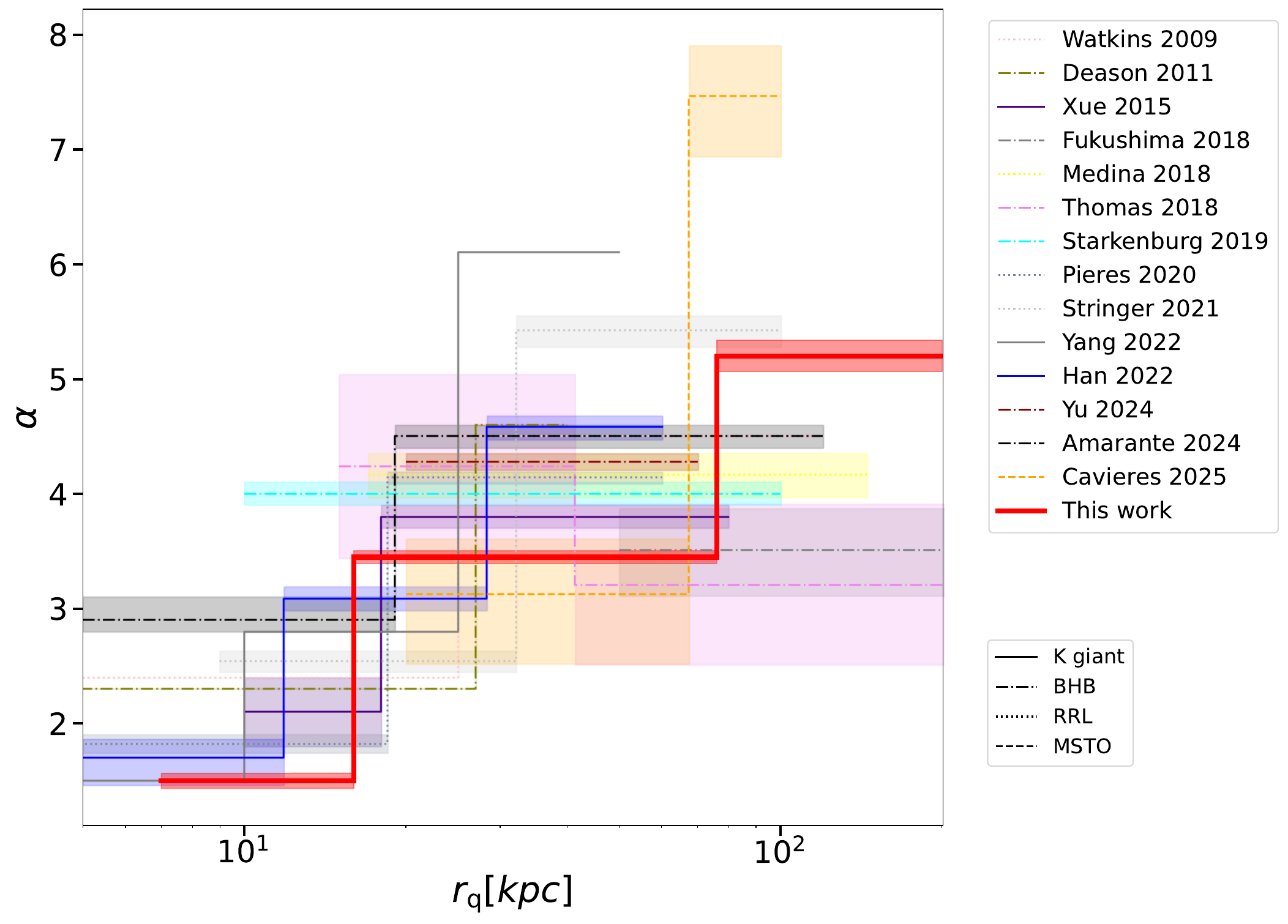}%
\end{center}
\caption{Radial density profiles of the MW stellar halo, comparing the result between this work (red solid line) and the literature. The vertical lines represent the break radii, and the horizontal lines denote the power indices within a given distance range, with the shaded range showing the uncertainties of power indices. The different line styles stand for various tracers.}
\label{fig:literature_comparison}
\end{figure}

Figure~\ref{fig:literature_comparison} shows the different radial density profiles from this work (red solid line) to those in the literature. The vertical lines represent the best-fit break radii. The horizontal lines stand for the power indices of the radial profiles, with the shaded range presenting the uncertainties. Many works adopt the double-power model, and the break radii vary from 10~kpc to 30~kpc. We adopt the interpretation from \cite{2021ApJ...923...92N} and \cite{Han_stellar_halo_density_profile}, and interpret the smaller break radii to be correlated with the apocenter of closer passage of GSE and the greater ones to be associated with the apocenter of the farther passage of GSE. The disparities in measured break radii could be explained by different footprints of surveys, as we have shown in Section~\ref{sec:Anisotropic stellar halo density distribution and the LMC wake} Table~\ref{table:regions}, the break radii vary with sky positions. 

We suggest that the break radius beyond 60~kpc is associated with the infall of the LMC (also see Section~\ref{sec:Anisotropic stellar halo density distribution and the LMC wake}). Most previous works failed to find such a great break radius, which is mainly due to the flux limit. For example, the limiting magnitude of the LAMOST survey in {\it Gaia} $G$-band is approximately 18~mag \citep{2015RAA....15.1095L}. \cite{2022AJ....164..241Y} only mapped the density distribution of the stellar halo within $\sim$50~kpc with LAMOST K giant. And for the H3 survey, the limiting magnitude in {\it Gaia} $G$-band is also approximately 18~mag \citep{2019ApJ...883..107C}, and thus \cite{Han_stellar_halo_density_profile} did not report a break radius at 60-70~kpc.

There can also be large differences in power-law indices, which are likely due to variations in sky positions and the presence of substructures \citep[e.g.][]{2015MNRAS.446.2274L}. Table~\ref{table:regions} shows that the first power indices can span from 0.5 to 2, the second vary from 2.6 to 3.9, and the third range from 4.8 to 7.5, reflecting significant anisotropies of the stellar halo. The anisotropic stellar halo can be explained by the remnants of major merger events in the sky. N-body simulations of the GSE merger \citep{2021ApJ...923...92N} and the ongoing infall of the LMC \citep{2019ApJ...884...51G,2021Natur.592..534C,2024MNRAS.534.2694S,2025arXiv250703663S} both demonstrate that two massive accretion events are sufficient to generate a highly anisotropic stellar halo, with position-dependent break radii, power-law slopes, and large-scale north–south asymmetry.
We refer the reader to \cite{joao}, which presents a more detailed discussion on the position dependence of power indices.

\section{Conclusions}
\label{sec:concl}

Based on the DR2 of the DESI MW Survey, we select a halo K giant sample ranging $8<r_\mathrm{GC}<200$kpc and perform detailed studies on the shape, orientation, radial density profile, and spatial anisotropies in density distribution of the MW stellar halo, which includes local overdensities and the LMC density wake of our MW stellar halo. 

First, a forward triaxial ellipsoidal model with triple power-law functional form is fit to the observed spatial distribution of halo K giants after convolving with the survey selection functions. The best-fit model reveals two break radii at 16 and 76~kpc. We interpret the smaller break radius to be associated with the first pericentric passage of GSE and the larger break radius to be correlated with the infall of LMC.
The stellar halo major axis is tilted by 44$\degree$ off the Galactic plane and 27$\degree$ away from the Sun-Galactic center axis. With the same model, we find the axis ratios are $1:p:q=10:8:7$, with the intermediate and minor axes lengths closer to each other than the major axis. 

By applying the forward model to halo K giants with different metallicities, we find that more metal-poor halo K giants have more extended radial profiles. 
We then allow the axis ratios and the major axis orientation to vary with galactocentric distances, $r_\mathrm{GC}$. Interestingly, the inner stellar halo at $r_\mathrm{GC}<\sim 30$~kpc is more oblate and more aligned with the disk, whereas the more distant stellar halo becomes more prolate and the major axis is more perpendicular to the MW disk. This indicates that the MW disk and inner stellar halo are misaligned with the outer stellar halo, while the outer stellar halo is more aligned with the Vast Polar Structure of 11 classical MW satellite galaxies. Numerical simulations have predicted the change in the direction of angular momentum of galaxy disks and inner halo, due to the infall of one or two massive satellites \citep[e.g.][]{2017MNRAS.472.3722G}, which may explain the twisted orientation and shape of the MW stellar halo in our measurement.
% Our results may suggest the flip in orientation of MW disc due to the infalling of Sagittarius, thus providing intriguing insights into the ``satellite plane'' problem for $\Lambda$CDM. 

Our MW stellar halo is full of substructures. We have successfully identified the HAC-N, HAC-S, and VOD overdensities associated with GSE. We identify a break radius at about 15~kpc in VOD and about 30~kpc in the HAC-N/S regions, consistent with \cite{Han_stellar_halo_density_profile}. 

The perturbation of the LMC to the MW stellar halo is also investigated. The transient LMC density wake at larger distances due to dynamical friction is identified in the Pisces region, with a second break radius at $\sim$60~kpc. For fields free of local overdensities, we identify that the footprint in the northern Galactic cap is about 4.5 times more overdense than the footprint in the southern cap between 50 and 100~kpc. And there is a break radius at $\sim$100~kpc and an overdense peak at 90~kpc in the northern Galactic cap, which corresponds to a detection of the collective density wake of the LMC. 

The DESI data have allowed us to explore the stellar halo in greater depth and detail than ever before. Our findings show that the stellar halo is littered with substructures, and we are able to model and relate these overdensities to known accretion events, and explain the varying shape and density of halo stars across the sky. With the full five-year DESI sample, we anticipate a factor-of-two reduction in the uncertainties of halo shape and orientation parameters, enabling a more precise measurement of the stellar halo twist signature and a better understanding of its origin.

The measurements presented in this paper can be accessed at \url{https://doi.org/10.5281/zenodo.17668310}, which contains all data points for the figures presented in this work.

\acknowledgments

This work is supported by NSFC (12573022, 12595312, 12273021), the National Key R\&D Program of China (2023YFA1605600, 2023YFA1605601), 111 project (No.\ B20019), and the Office of Science and Technology, Shanghai Municipal Government (grant Nos. 24DX1400100, ZJ2023-ZD-001). We thank the sponsorship from Yangyang Development Fund. The computations of this work are carried on the National Energy Research Scientific Computing Center (NERSC) and the Gravity supercomputer at the Department of Astronomy, Shanghai Jiao Tong University. SK acknowledges support from the Science \& Technology Facilities Council (STFC) grant ST/Y001001/1. MV  acknowledges support from NASA ATP award (80NSSC20K0509). APC acknowledges support from the Taiwan Ministry of Education Yushan Fellowship, MOE-113-YSFMS-0002-001-P2, and Taiwanese National Science and Technology Council grant 112-2112-M-007-009. LBeS acknowledges support from CNPq (Brazil) through a research productivity fellowship, grant no. [304873/2025-0].

We are grateful for Martin White to be our DESI publication handler. WW is grateful for useful discussions with Yanjun Sheng, Jianhui Lian, Baitian Tang, Jie Wang, Zhaozhou Li, Ling Zhu, Enci Wang, Hongxin Zhang, Yu Rong, Huiyuan Wang and Bojun Tao. 

This material is based upon work supported by the U.S. Department of Energy (DOE), Office of Science, Office of High-Energy Physics, under Contract No. DE–AC02–05CH11231, and by the National Energy Research Scientific Computing Center, a DOE Office of Science User Facility under the same contract. Additional support for DESI was provided by the U.S. National Science Foundation (NSF), Division of Astronomical Sciences under Contract No. AST-0950945 to the NSF’s National Optical-Infrared Astronomy Research Laboratory; the Science and Technology Facilities Council of the United Kingdom; the Gordon and Betty Moore Foundation; the Heising-Simons Foundation; the French Alternative Energies and Atomic Energy Commission (CEA); the National Council of Humanities, Science and Technology of Mexico (CONAHCYT); the Ministry of Science, Innovation and Universities of Spain (MICIU/AEI/10.13039/501100011033), and by the DESI Member Institutions: \url{https://www.desi.lbl.gov/collaborating-institutions}. Any opinions, findings, and conclusions or recommendations expressed in this material are those of the author(s) and do not necessarily reflect the views of the U. S. National Science Foundation, the U. S. Department of Energy, or any of the listed funding agencies.

The authors are honored to be permitted to conduct scientific research on I'oligam Du'ag (Kitt Peak), a mountain with particular significance to the Tohono O’odham Nation.

For the purpose of open access, the author has applied a Creative
Commons Attribution (CC BY) licence to any Author Accepted
Manuscript version arising from this submission. % Added by Sergey Koposov, that's my university 

We thank the anonymous referee for his/her time and effort spent on reading and comment our paper.

%\clearpage

\appendix
\section{Effect of distance uncertainties}
\label{app:error_on_model}

In this appendix, we briefly discuss how the distance uncertainties of K giants could change the radial density profile and hence impact our results. 
We generate a mock density distribution with a triple power-law radial density profile without any observational uncertainties, which covers the whole sky (hereafter, we name it the ideal profile). We assume the parameters of the ideal profile are our best-fit values, as shown in Figure~\ref{fig:mcmc_2rb}. We then perturb the ideal density distribution with the same distance error from the observation, and then further convolve the perturbed density distribution with the DESI footprint and the angular $+$ radial selection functions. We run MCMC again to get the best-fit parameters of the perturbed density distribution, as for real data. 

\begin{figure}
\begin{center}
\includegraphics[width=0.4\textwidth]{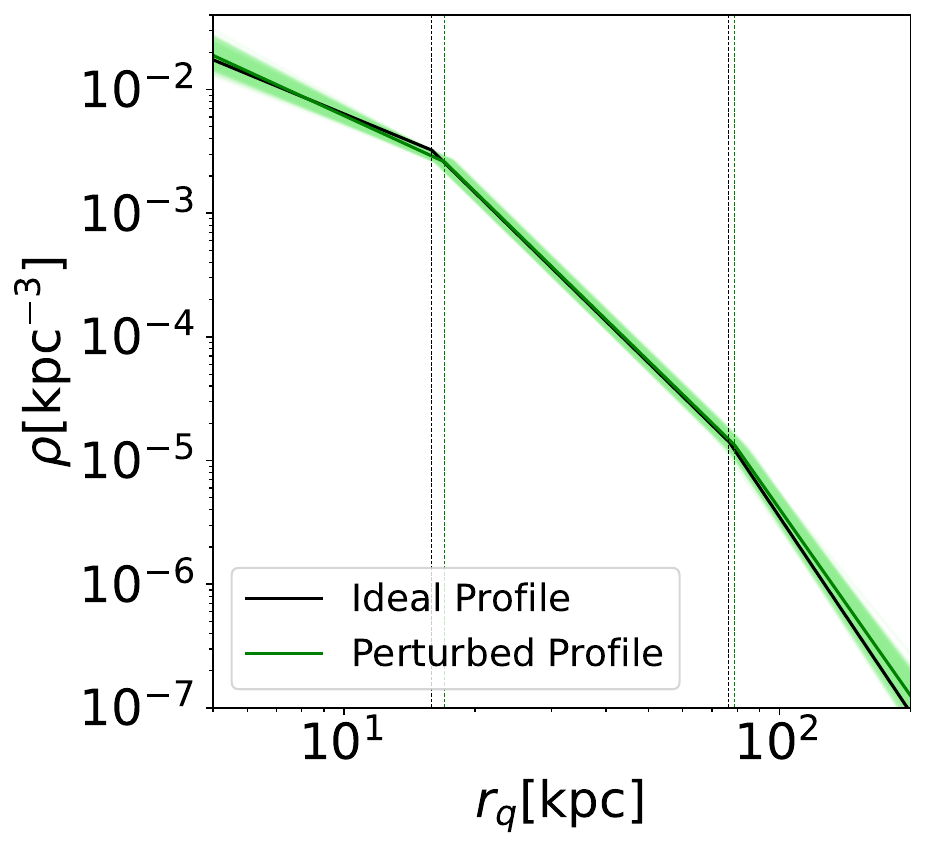}%
\end{center}
\caption{The black solid line refers to the selection effect free density profile based on a mock realization (ideal profile) of stars. The green line is the selection effect free best-fit model for the ideal profile perturbed with observational error. The green shaded region shows the statistical uncertainties of the best-fit model. The vertical dashed lines stand for the break radii of the original profile (black) and the recovered profile (green) after perturbation. Both profiles are renormalized to unity.}
\label{fig:perturbed_2rb_compare}
\end{figure}

Figure~\ref{fig:perturbed_2rb_compare} shows two radial density profiles of the ideal profile and the best-fit profile of the perturbed density distribution, with the green shaded region representing the model uncertainties of the perturbed density profile. The perturbed density profile shows larger break radii and different power indices. However, the deviation between the ideal profile and the perturbed profile is still within the model uncertainties. We have repeated this process multiple times to get different realizations of the perturbed profile. For all realizations, the difference is smaller than the statistical error. Hence, we believe that the distance uncertainty is not prominent in this work. 

\bibliography{master}{}
\bibliographystyle{aasjournal}

%\bibliography{master}{}
%\bibliographystyle{aasjournal}

%% This command is needed to show the entire author+affiliation list when
%% the collaboration and author truncation commands are used. It has to
%% go at the end of the manuscript.
%\allauthors

%% Include this line if you are using the \added, \replaced, \deleted
%% commands to see a summary list of all changes at the end of the article.
%\listofchanges

\end{document}